\pdfoutput=1
\documentclass[a4paper,twocolumn]{article}          
\usepackage{graphicx}
\usepackage{colortbl}
\usepackage{url}
\usepackage{amssymb,amsfonts,amsmath}
\usepackage{algorithm}
\usepackage{algpseudocode}
\usepackage[misc,geometry]{ifsym} 
\usepackage[colorlinks=true,breaklinks]{hyperref}
\usepackage[round,sort&compress]{natbib}
    \usepackage{sectsty}
    \sectionfont{\normalsize}
    \subsectionfont{\normalsize}
    
    \usepackage[
    top    = 2.50cm,
    bottom = 2.50cm,
    left   = 2.75cm,
    right  = 2.75cm]{geometry}

 \hypersetup{
   colorlinks,
   citecolor=Violet,
   linkcolor=Red,
   urlcolor=Blue}

\ifx\pdfoutput\@undefined\usepackage[usenames,dvips]{xcolor}
\else\usepackage[usenames,dvipsnames]{xcolor}
\IfFileExists{pdfcolmk.sty}{\usepackage{pdfcolmk}}{} 
\fi

\title{Exploring Programmable Self-Assembly in Non-DNA based Molecular Computing}

\author{Germ\'an Terrazas\thanks{Institute of Advanced Manufacturing, Faculty of Engineering, University of Nottingham, UK. E-mail: German.Terrazas@nottingham.ac.uk} 
\and Hector Zenil \thanks{Kroto Research Institute, Behavioural and Evolutionary Theory Lab, Department of Computer Science University of Sheffield, UK. E-mail: hectorz@labores.eu}
\and Natalio Krasnogor \thanks{School of Computing Science and Centre for Bacterial Cell Biology, Newcastle University, UK. E-mail: Natalio.Krasnogor@newcastle.ac.uk (\Letter)}
}

\date{\footnotesize The final publication is available at link.springer.com DOI: 10.1007/s11047-013-9397-2}
\begin{document}

\maketitle

\begin{abstract}
Self-assembly is a phenomenon observed in nature at all scales where autonomous entities build complex structures, without external influences nor centralised master plan. Modelling such entities and programming correct interactions among them is crucial for controlling the manufacture of desired complex structures at the molecular and supramolecular scale. This work focuses on a programmability model for non DNA-based molecules and complex behaviour analysis of their self-assembled conformations. In particular, we look into modelling, programming and simulation of porphyrin molecules self-assembly and apply Kolgomorov complexity-based techniques to classify and assess simulation results in terms of information content. The analysis focuses on phase transition, clustering, variability and parameter discovery which as a whole pave the way to the notion of complex systems programmability.
\end{abstract}

\section{Introduction}
Self-assembly research and practice \citep{KraGusPelVer06} (regardless of the scale at which it operates) often encounters three key problems (a) the forward problem, (b) the backward problem (also known as the designability problem) and (c) the yield problem \citep{pelesko2007}. The forward problem is concerned with trying to predict what the final product of the self-assembly process would be, given a set of objects, environmental conditions and the natural laws (physical, chemical, biological) that are prevalent at a given specific scale. Usually the forward problem is addressed through the use of simulations and mathematical models. The backward problem, the most difficult of the three, addresses the issue of how the objects and the environment that contains them can be designed in such a way that the final outcome of the self-assembling process is a specific pre-ordained one. As surveyed in \cite{pelesko2007}, this problem is usually addressed through very sophisticated heuristics methods as, in lieu of the NP-hardness (in some cases even undecidability) of the most relevant backward problems, exact analytical solutions are very rarely achievable. The third problem, that of the yield of a self-assembly process, is related to the estimation and control of how many of the intended target self-assembled objects one can expect from a particular self-assembling system (this problem is ubiquitous in the chemical sciences). The observation that ``self-assembly and computation are linked by the study of mathematical tiling'' \citep{Rothemund01022000} has produced a step change in the way the forward, yield and, more significantly, the backward problems in molecular self-assembly are dealt with. More specifically, \cite{winfree1996computational,winfree1998design,mao2000logical,soloveichik2005computational} have shown that universal computation can be carried out by self-assembling discrete DNA tiles in a 2D plane and, by utilizing the power of universal computation, complex DNA-based patterns have been implemented in the lab through a clever programming of the DNA tiles. Indeed, linking self-assembly and computation provides a powerful new approach to addressing profound questions about the controllability of complex physico-chemical nanosystems. We could ask, for example \citep{rothemund2000program,adleman2001running,AdlemanCGHKER02,soloveichik2005computational} {\it what are the least complex molecular tiling 'motifs' which may be exploited in the programming of self-assembly 2D lattices with specific geometries?} It was shown in \cite{adleman2001running,AdlemanCGHKER02} that answering this question might, in some cases, be a computationally undecidable query while in other cases it might give rise to NP-hard problems, thus it is strongly suspected that exact polynomial time deterministic algorithms do not exist for these problems. It is important to remark that the idea is not necessarily to use self-assembly for computational purposes (as in DNA computing) but, rather, the other way around: to use computation in such a way as to program nano tiles so they self-assemble, with exquisite detail, into specific patterns. That is, computation embedded in the tiles' design allows for an enhanced control of the self-assembling entities; this remarkable formal connection between self-assembly and computation is the subject of our work.

We have demonstrated that a combination of experiments, modelling and evolutionary computation can automatically program idealized models of discrete self-assembly tiling systems \citep{terrazas2005,terrazas2007} and also self-organising gold nanoparticle assemblies \citep{SieMarVanMorKra2007} in such a way that they achieve specific self-assembled conformations. In one of our studies we concentrated on a system that consisted of so called Wang tiles. Wang tiles "live" in a 2-dimensional world and can freely move in this 2D space. When two tiles collide, the glue type of the colliding sides is used to decide whether the tiles should stick to each other or bounce back. Given a set of glue types with their characteristic strengths and a given temperature, we were able to solve the backward problem and provide answers to the question of {\it what is the (optimal) family of tiles that will self-assemble into a specific spatio-temporal pattern?} We have also shown \citep{terrazas2005,Siepmann06,SieMarVanMorKra2007,TerrazasSKK07} that it is possible to evolve the parameters of a cellular automata-based Monte Carlo model to coerce a specific spatio-temporal pattern closely matching observed nanoscience experiments imaged with an atomic force microscope while substantially speeding-up the process of nanoscanning \citep{woolley2011automated}. In contrast to work mentioned previously, here we focus on extensive simulations of non DNA-based molecules deposited on a suitably processed solid substrate and on complex behaviour analysis. Indeed, outside DNA-based systems (\cite{rothemund2000program,winfree2004proofreading,soloveichik2005computational}), analogues of programmable molecular tiling of complex self-assembling patterns have yet to be systematically studied and this paper is a first step in that direction. For a survey of self-assembly systems at various scales and under various physical embodiments please refer to \cite{KraGusPelVer06}.

Different approaches have been employed for the characterisation, quantification and classification of complex behaviour. Kolmogorov complexity \citep{Kolmogorov1965QuantitativeInformation} is the mathematical measure of randomness that together with some metric variations is well equipped to tell apart structure from simplicity as a measure of information. The Kolmogorov complexity of a given object is defined as the length of the shortest program for computing such object by a universal Turing machine \citep{Chaitin69onthe}. In other words, this method characterises the complexity of an object by the length of its shortest description, that is, the minimum number of symbols needed for a computer program to reproduce such object. Kolmogorov complexity is an uncomputable function and one of its approximations is implemented by lossless compression algorithms. Such approximation has been applied to study the qualitative dynamical properties of cellular automata \citep{zenil10compressionbased}, classification of cellular automata \citep{Duba01a}, classification of biological sequences \citep{FerraginaGGMV07} and data mining \citep{keogh07compression}, to name but a few. An important result of Kolmogorov complexity is the {\it Normalised Compression Distance} \citep{Li2004} which is a measure of similarity between two given objects in terms of information.  The information distance between two objects is defined as the amount of information required to compute one object given the other. This metric has been applied to different purposes in a wide range of research fields such as pattern recognition, data mining, clustering, evolutionary design and classification \citep{CilibrasiV05,Siepmann06,TerrazasSKK07,Vitanyi12}, and also employed as a base for defining information distance between multiple objects \citep{Vitanyi11}.

In this work, we introduce a simple mathematical model that captures relevant porphyrin molecules dynamics and blends with a classic stochastic algorithm into a self-assembly simulation system. The behaviour of this system is mainly governed by programmable porphyrin molecules, the different instances of which give rise to an extensive variety of morphologically complex self-assembled structures where some of these closely match supramolecular conformations observed in porphyrin molecules deposited onto a $A(111)$ solid processed gold substrate. Our aim is to characterise qualitative traits of the complex behaviour captured by our system, i.e. the resulting self-assembled aggregates, in terms of Kolmogorov complexity and information distance. Next section introduces the porphyrin molecules programmability model followed by a full description of our self-assembly simulation system, description of experiments and simulation results. Then, the characterisation of the system follows focusing on phase transition, clustering, variability, parameter discovery, orthogonality and finishes with an introduction to the notion of complex system programmability.
\begin{figure*}[ht!]
\centering
\begin{tabular}{cc}
\includegraphics[scale=0.5]{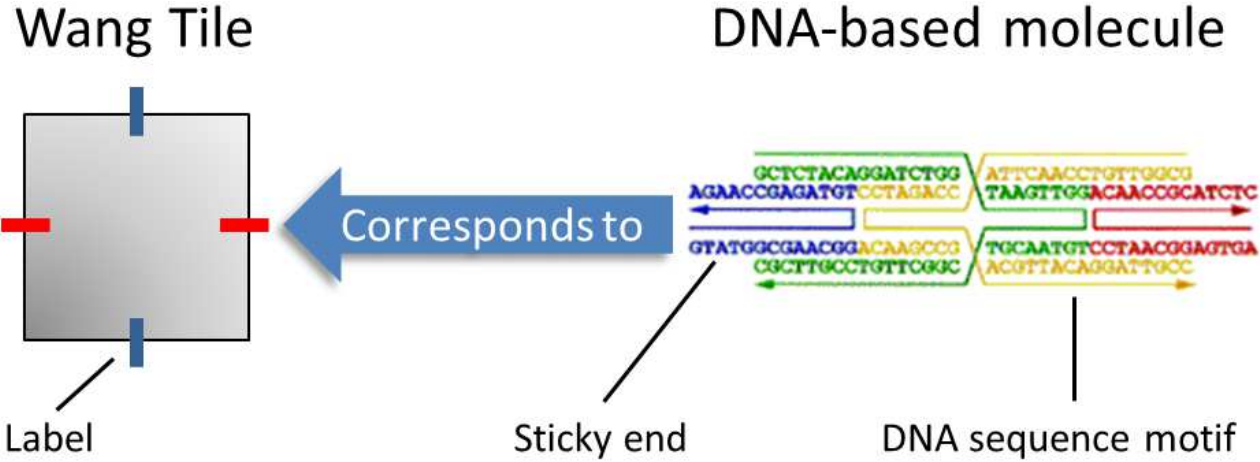} & \includegraphics[scale=0.6]{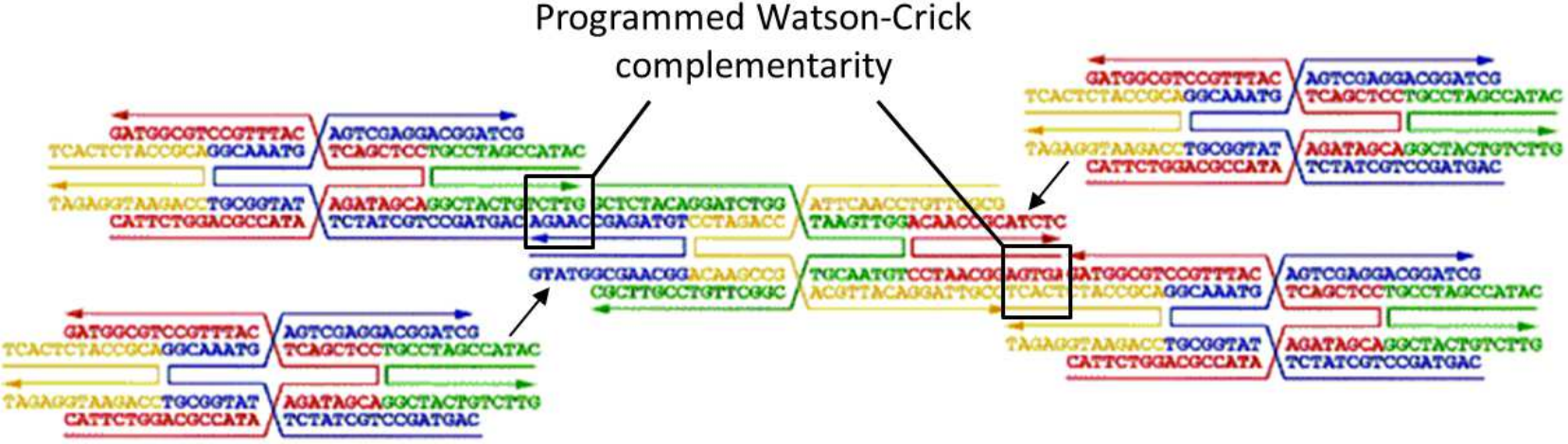}\\
(a) & (b)
\end{tabular}
\caption{\label{winfree_embodies} \small Correspondence between a Wang tile and a double cross-over, antiparallel, odd-spacing (DAO) molecule where tile labels map to sticky ends of unique DNA sequence (a). Self-assembly of  molecular units into two-dimensional lattice by means of programmed Watson-Crick complementary sticky ends (b). (Reported in \cite{winfree1998design})}
\end{figure*}

\section{Programmability Model}

The programmability of detailed structure of matter at nano scale has been reported in \cite{winfree1998design} where the aim is to design molecular DNA-based units with predictable and controllable interactions that self-assemble into two-component lattices, with a stripe every other unit, and into four-component lattices, with a stripe every fourth unit. In order to achieve this, the mathematical theory behind Wang tiles \citep{wang1961} has been employed for the physical and operational design of antiparallel double-crossover (DX) DNA-based  molecules which act as molecular tiles with programmable interactions (see Fig. \ref{winfree_embodies} (a)). This resulted in the production of $12.6nm$ wide double cross-over, antiparallel, odd-spacing (DAO) molecules and double cross-over, antiparallel, even spacing (DAE) molecules, the corners of which consist of single stranded sticky ends of unique DNA sequence. Correct association among DX units is then achieved by carefully programming their sticky ends with Watson-Crick complementarity (see Fig. \ref{winfree_embodies} (b)) in such a way that undesired associations are unlikely to take place. For such purpose, the principle of sequence symmetry was employed in order to maximise the free energy difference between desired and alternative conformations.

\begin{figure}[h]
\centering
\begin{tabular}{c}
\includegraphics[scale=0.60]{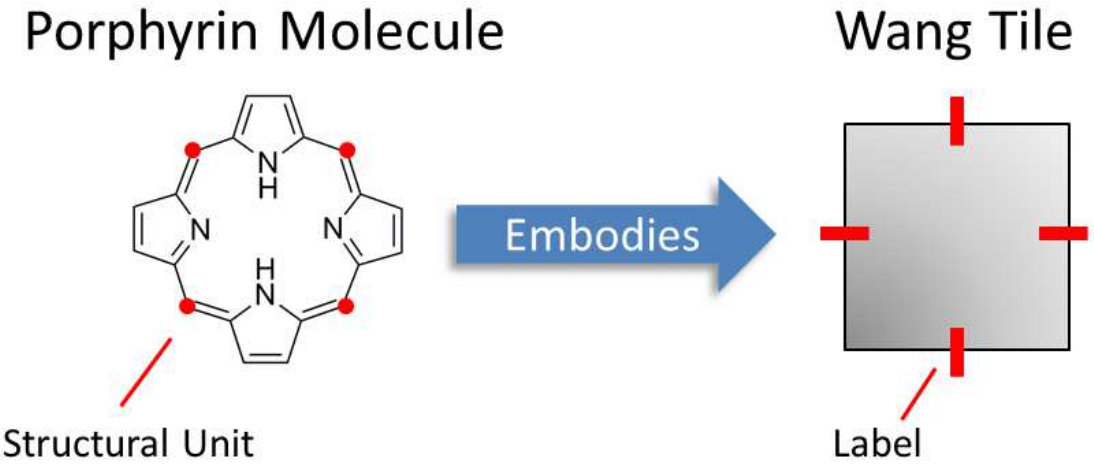}\\
[2ex]
\includegraphics[scale=0.60]{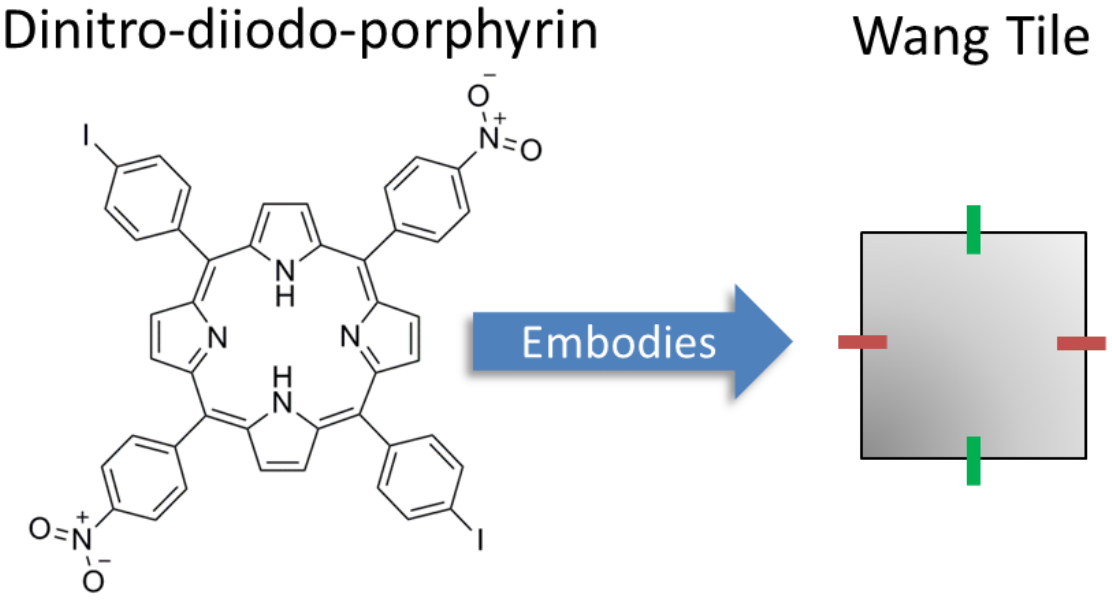}
\end{tabular}
\caption{\label{embodies} Wang tile as embodiment of a porphyrin with labels corresponding to molecular structural units (top) set with nitrogen and iodine originate a functionalised porphyrin molecule embodied by a Wang tile where different labels correspond to functional groups (bottom).}
\end{figure}

In our work, we employed porphyrin molecules which are planar molecular units with a dimension of $2.89nm^2$, fourfold symmetry and suitable for solid substrate deposition. The chemical structure of a porphyrin molecule reveals four structural units which can be synthesised with substituent functional groups, hence giving as a result {\it functionalised (programmable) porphyrin molecules}. The intermolecular hydrogen bonding and van der Waals interactions among such substituents allow diverse self-assembly complexity together with a high degree of reversibility and highly dynamic pattern formation. We have currently synthesised porphyrins with iodine, carboxylic acid, pyridine, bromine and nitro functional groups. Their chemical structures as well as some of the currently estimated relative binding strengths between them are collected in Table \ref{relative_strengths}. 
\begin{table}[h]
\centering
\small\addtolength{\tabcolsep}{-3.5pt}
\begin{tabular}{lccccc}
& \includegraphics[scale=0.12]{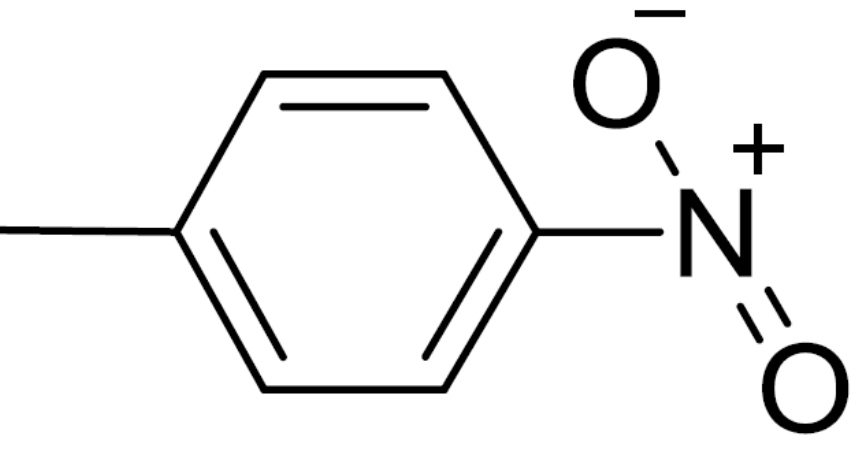} & \includegraphics[scale=0.12]{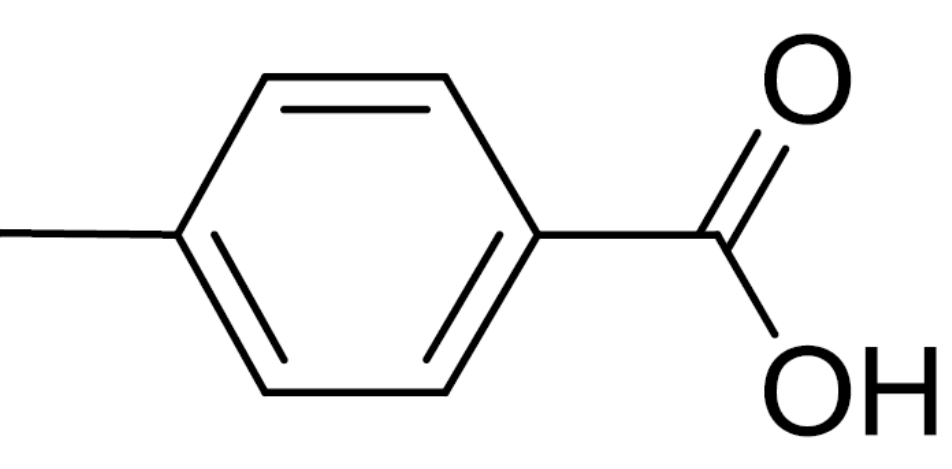} & \includegraphics[scale=0.12]{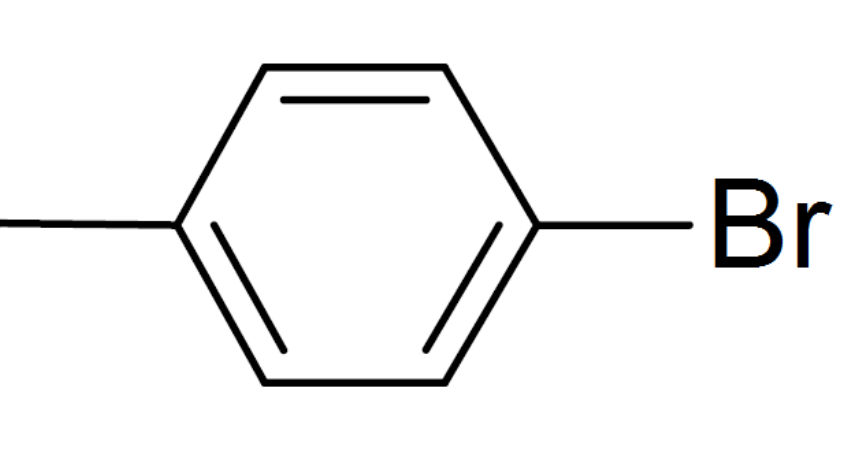} & \includegraphics[scale=0.12]{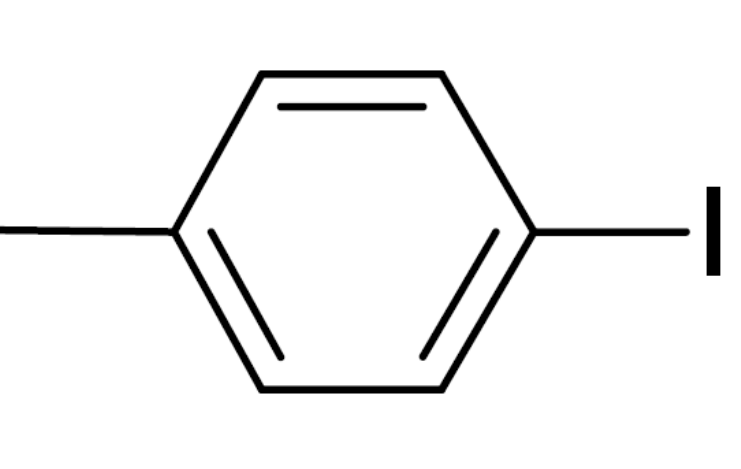} & \includegraphics[scale=0.12]{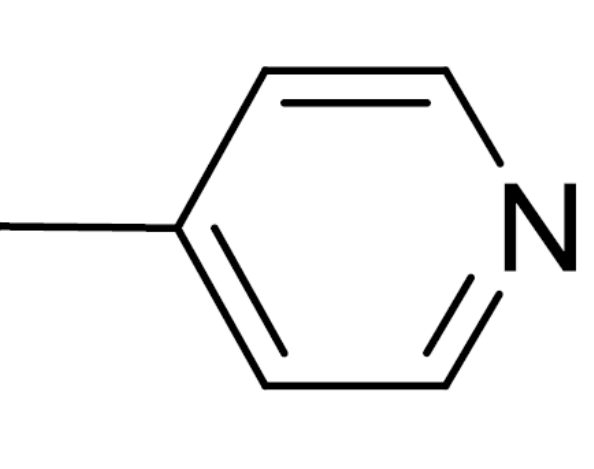}\\
[1.0ex]
\includegraphics[scale=0.12]{h_nitro.pdf}  & \raisebox{2.1ex}{N/A} & \raisebox{2.1ex}{N/A}  & \raisebox{2.1ex}{N/A}  & \raisebox{2.1ex}{0.13} & \raisebox{2.1ex}{N/A}\\
[1.0ex]
\includegraphics[scale=0.12]{h_carbox.pdf}   &\cellcolor[gray]{0.9}& \raisebox{2.1ex}{0.30} & \raisebox{2.1ex}{N/A} & \raisebox{2.1ex}{N/A} & \raisebox{2.1ex}{0.39}\\
[1.0ex]
\includegraphics[scale=0.12]{h_bromine.pdf}  &\cellcolor[gray]{0.9}&\cellcolor[gray]{0.9}& \raisebox{2.1ex}{1.00} & \raisebox{2.1ex}{N/A} & \raisebox{2.1ex}{N/A}\\
[1.0ex]
\includegraphics[scale=0.12]{h_iodine.pdf}   &\cellcolor[gray]{0.9}&\cellcolor[gray]{0.9}&\cellcolor[gray]{0.9}& \raisebox{2.1ex}{0.087} & \raisebox{2.1ex}{0.17}\\
[1.0ex]
\includegraphics[scale=0.12]{h_pyridine.pdf} &\cellcolor[gray]{0.9}&\cellcolor[gray]{0.9}&\cellcolor[gray]{0.9}&\cellcolor[gray]{0.9}& \raisebox{2.1ex}{0.10}\\
\end{tabular}
\caption{\label{relative_strengths} \small Functional groups employed to synthesise porphyrin molecules. The approximate binding energy values between nitro ($NO_2$), carboxylic acid ($O_2H$), bromine ($Br$), iodine ($I$) and pyridine ($N$) are expressed in electro volts ($\,eV$).}
\end{table}
Thus, the programmability of the structural units on porphyrin molecules is the point of interest in our work. The potential of our approach  is that by employing non-DNA based molecules one can access different chemical/physical systems and eventually embed computation in them. By being different than DNA, porphyrins based computation will be able to operate under temperature regimes, solvents, PH levels, concentrations, etc., unlike those required for DNA. Also,  porphyrin tiles are considerable smaller than DNA tiles. In addition, porphyrins self-assembly takes place not in bulk solution but rather parallel to a surface, hence naturally leading to a 2D self-assembly stratagem and thus contrasting to DNA-based strategies which employ highly complex 3D motifs for self-assembly even when dealing with 2D patterns. We show two example applications in \citealp[pp. 4-5]{TerLuiKras2013}. In the first one we demonstrate that counters \citep{cheng2004optimal,MoissetdeEspanes2008225} can be engineered with porphyrin tiles. These counters are employed to systematically self-assemble a two-dimensional structure in such a way that some porphyrin tiles form backbones directing the physical extent other porphyrin tiles can ``flood'' a well-defined region. Another example is provided of a two-state probabilistic automaton that can output a globally complex pattern made up of internally ordered substructures. Thus, we model a physico-chemical system where fully functionalised porphyrin molecules deposited onto a gold processed substrate perform intermolecular interactions which drive the creation of self-assembled supramolecular aggregates. We choose Wang tiles as physical embodiment since these are square in shape with labelled edges and undergo tile-to-tile interactions, hence exhibiting not only a morphological correspondence to functionalised porphyrin molecules, but also a functional mapping to the intermolecular interactions. An illustration of such correspondence between Wang tiles and porphyrin molecules is shown at the top of Fig. \ref{embodies}. From here onwards we refer to such embodiment as {\it porphyrin-tile}, which could be defined as either {\it iso-functionalised} when its four sides are programmed with the same functional group and as {\it hetero-functionalised} when its four sides are programmed with different functional groups. An example of a hetero-functionalised porphyrin molecule and its corresponding porphyrin-tile embodiment is depicted at the bottom of Fig. \ref{embodies}. In addition, the substrate where molecules are deposited and on which aggregates are formed is modelled as a two-dimensional square site lattice set with periodic boundary conditions where each position is occupied by only one porphyrin-tile at a time.

\section{Porphyrin-tiles Kinetic Monte Carlo System}

\begin{figure}[ht]
\centering
\begin{tabular}{cc}
\includegraphics[scale=0.2]{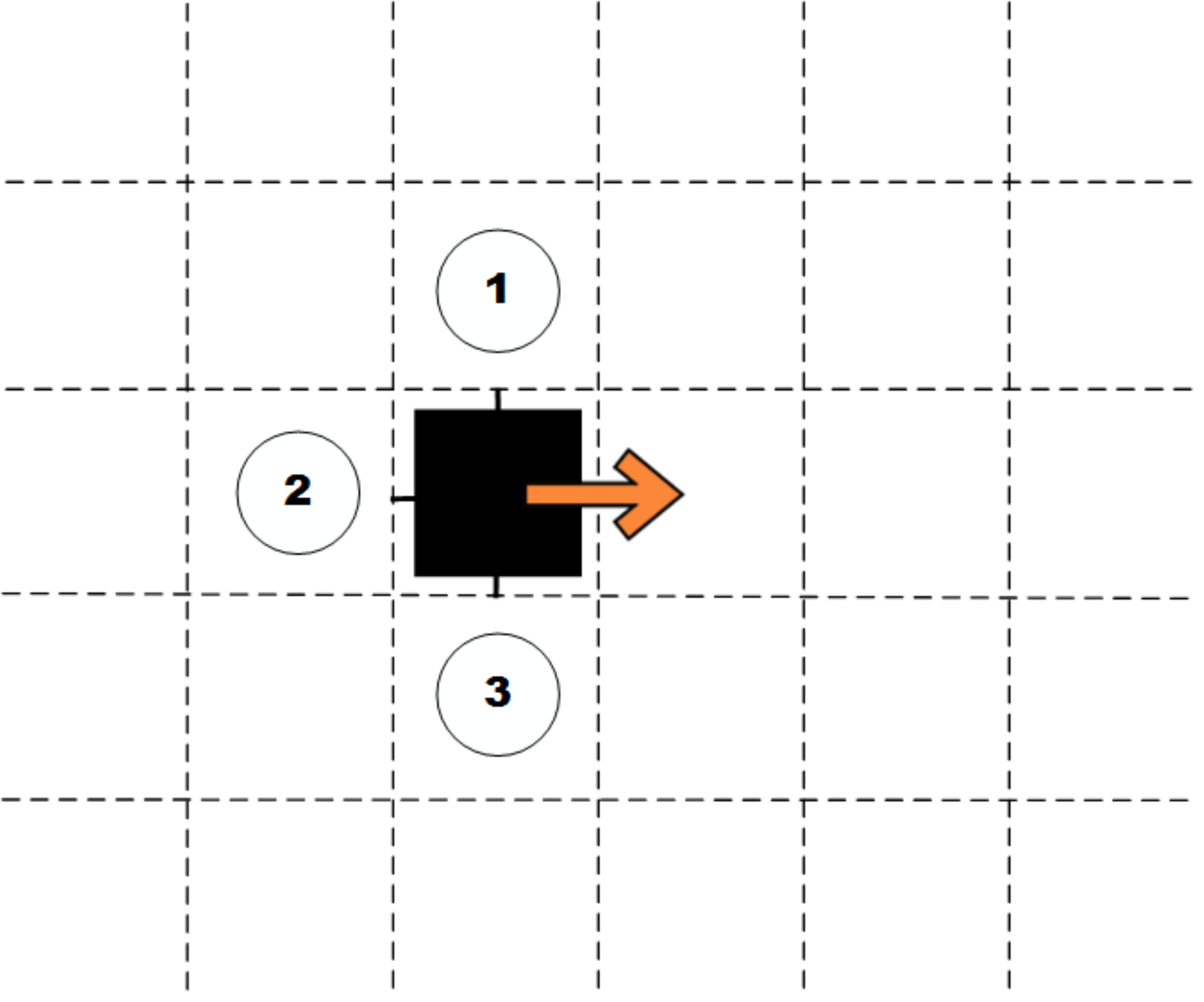} & \includegraphics[scale=0.19]{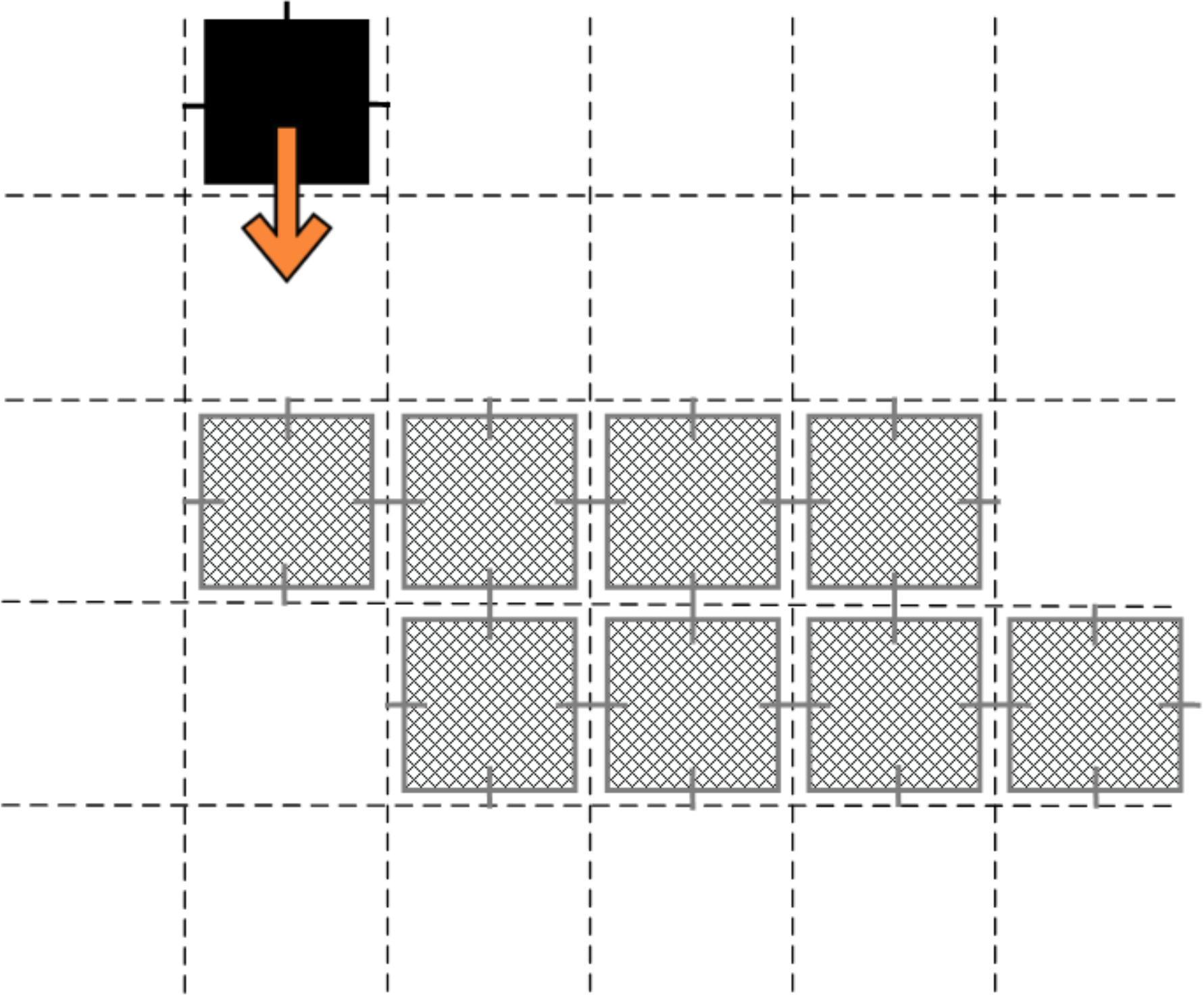}\\
(a) & (b)\\
[2ex]
\includegraphics[scale=0.2]{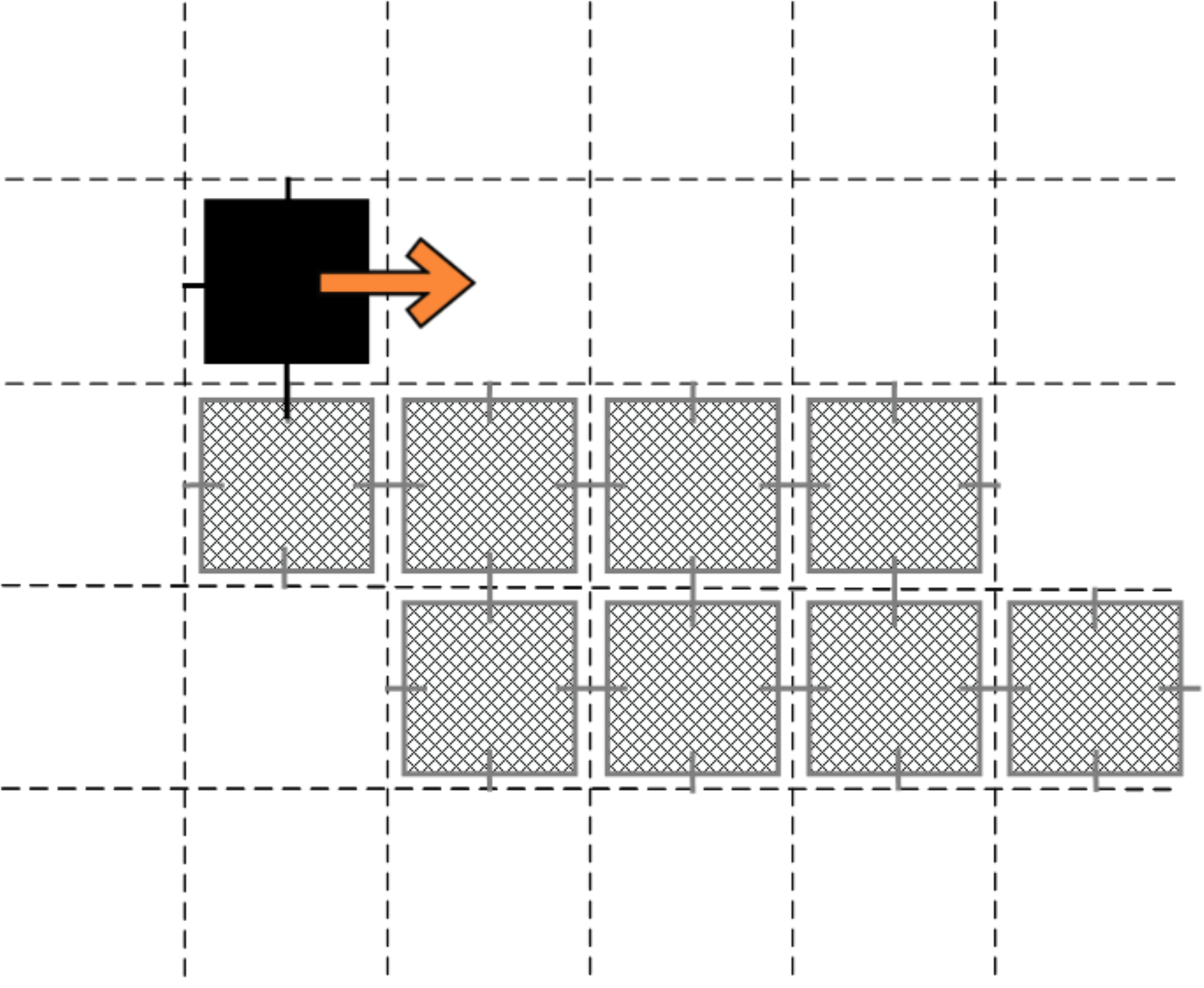} & \includegraphics[scale=0.19]{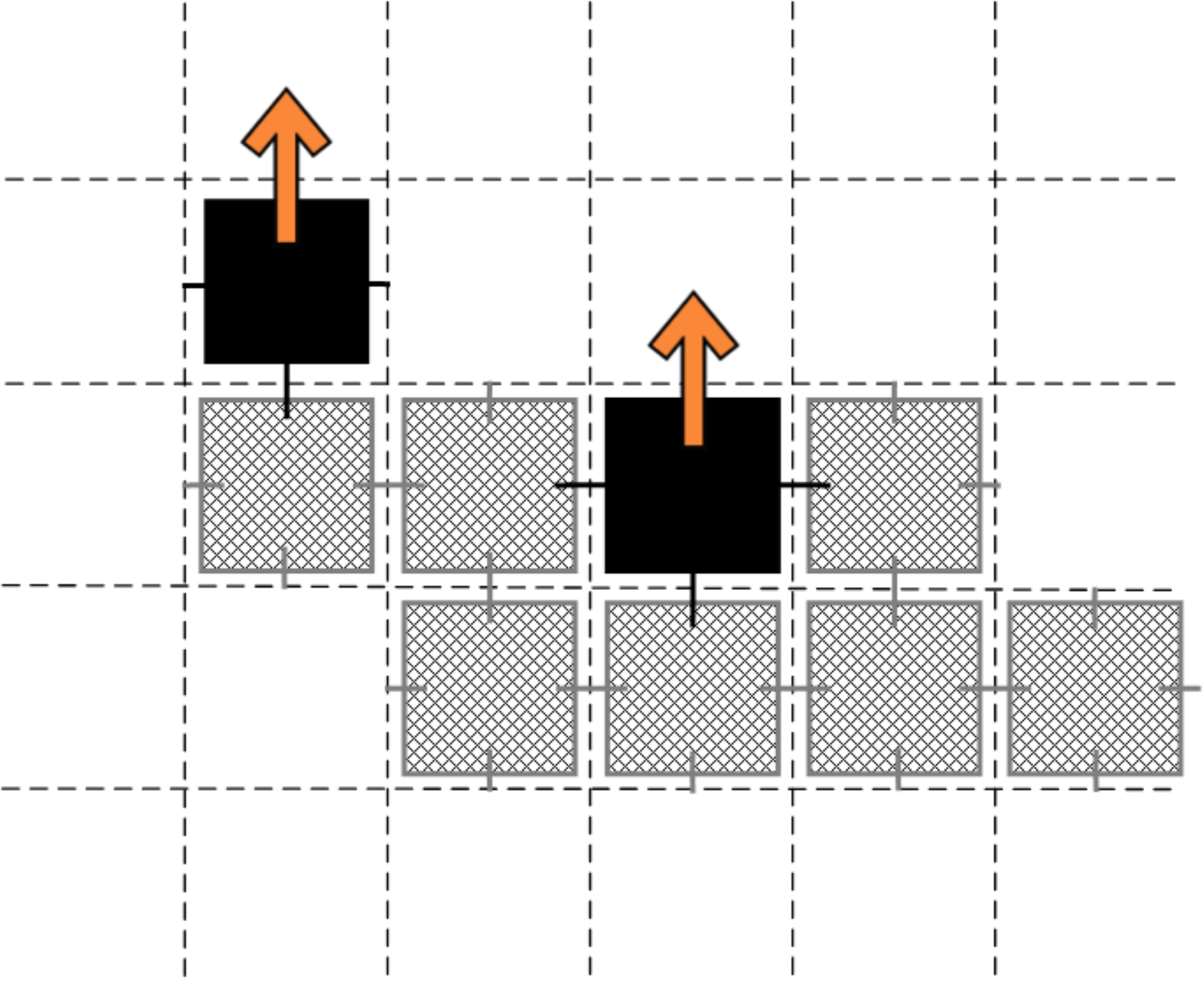}\\
(c) & (d)
\end{tabular}
\caption{\label{diffusion_motion} \small Symbolic examples of a molecule hop from position $(i,j)$ to position $(i, j+1)$ together with its neighbouring positions (a), diffusion of a molecule across the lattice without interacting with neighbouring molecules (b), diffusion of a molecule along an aggregate (c) and departure of a molecule from an aggregate (d).}
\end{figure}

The Monte Carlo family of methods are stochastic simulation algorithms used to model the behaviour of complex systems without the need to solve analytically the equations governing the system in question. These methods are generally good when a fast approximation of the overall behaviour of the system is needed \citep{PhysRevE.85.031907}. Inspired by a kinetic Monte Carlo (kMC) system programmed for the simulation of nucleation and growth of thin metal films onto amorphous substrates \citep{PhysRevB.55.7955}, we have designed and developed a {\it porphyrin-tiles kMC} system for the simulation of the self-assembly process between functionalised porphyrin molecules. Energy interactions among neighbouring molecules are at the core of the system dynamics and in our case the neighbourhood for a molecule at position $(i,j)$ is defined as von Neumann type. A symbolic example of a molecule hop from position $(i, j)$ to position $(i, j$+$1)$ together with its neighbouring positions is depicted in Fig. \ref{diffusion_motion} (a).
\\
Energy interactions among neighbouring molecules are exploited here for capturing three phenomena: {\it deposition}, {\it motion} and {\it rotation} of a molecule on the substrate. In particular, deposition models the arrival of a molecule onto an empty position of the substrate, i.e. the entrance of a porphyrin-tile to an unoccupied position $(i,j)$ of the lattice. Motion models the translation of a molecule to one of its four neighbouring empty positions of the substrate, i.e. the movement of a porphyrin-tile located at position $(i,j)$ into one of its four unoccupied nearest neighbouring positions $(i$+$1,j)$, $(i, j$+$1)$, $(i$-$1, j)$ or $(i, j$-$1)$ by considering three cases: the diffusion of a molecule across the lattice without interacting with neighbouring molecules as shown in Fig. \ref{diffusion_motion} (b), diffusion along an aggregate as depicted in Fig. \ref{diffusion_motion} (c) or departure of a molecule from an aggregate as illustrated in Fig. \ref{diffusion_motion} (d). Rotation models spinning of a molecule on its centre of mass, i.e. the $\pm 90$ degrees gyration of a porphyrin-tile on its geometrical midpoint.
\begin{figure}[h!]
\centering
\begin{tabular}{cc}
\includegraphics[scale=0.4]{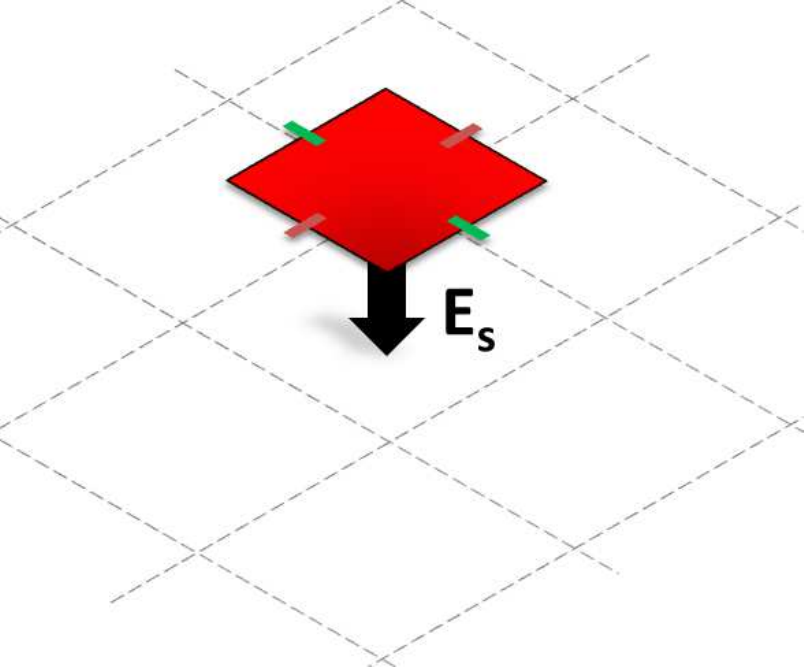} & \includegraphics[scale=0.3]{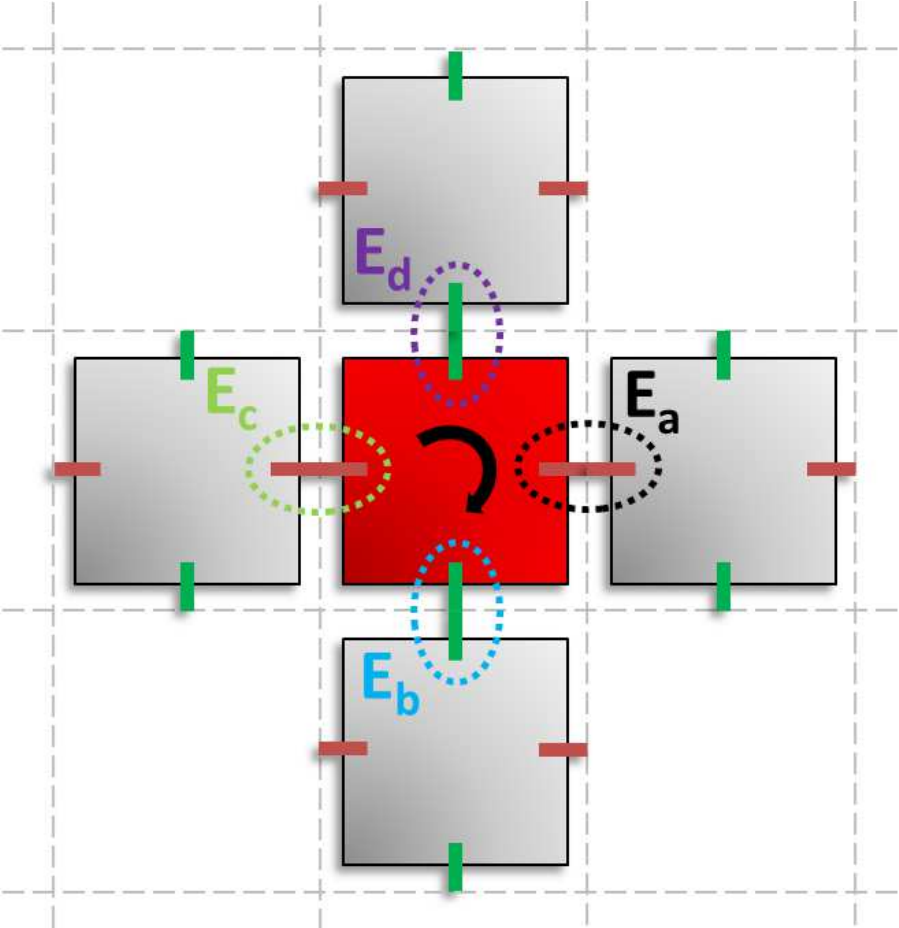}\\
(a) & (d)\\
[2ex]
\includegraphics[scale=0.3]{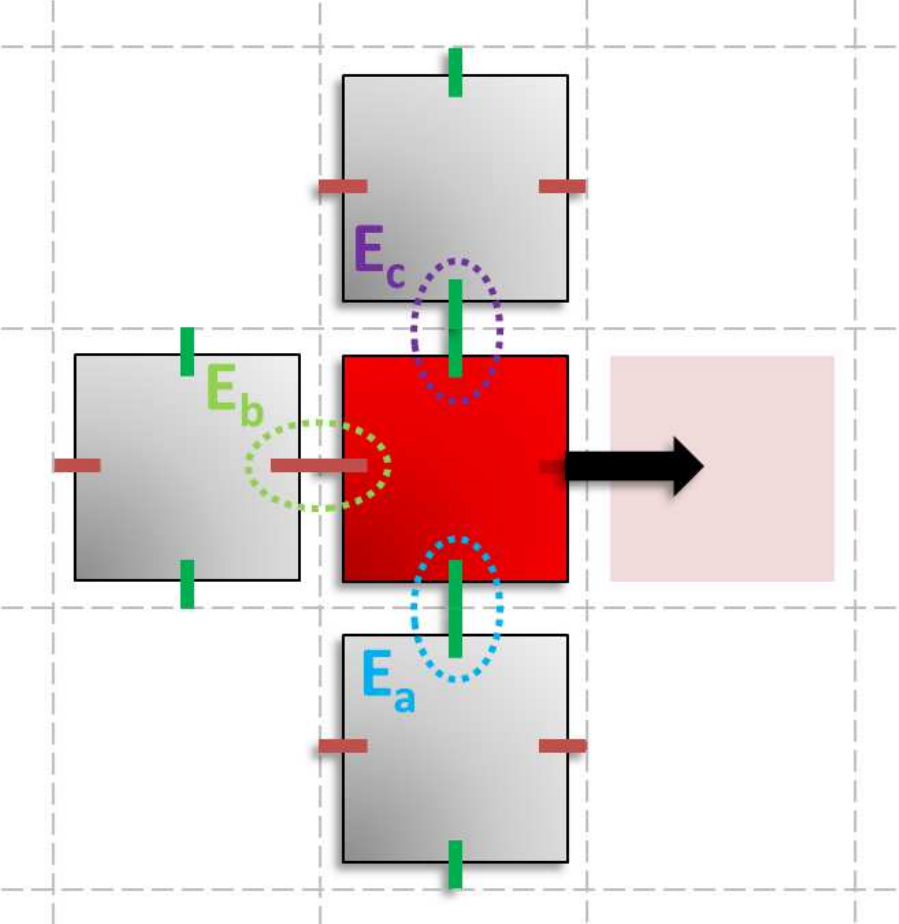} & \includegraphics[scale=0.3]{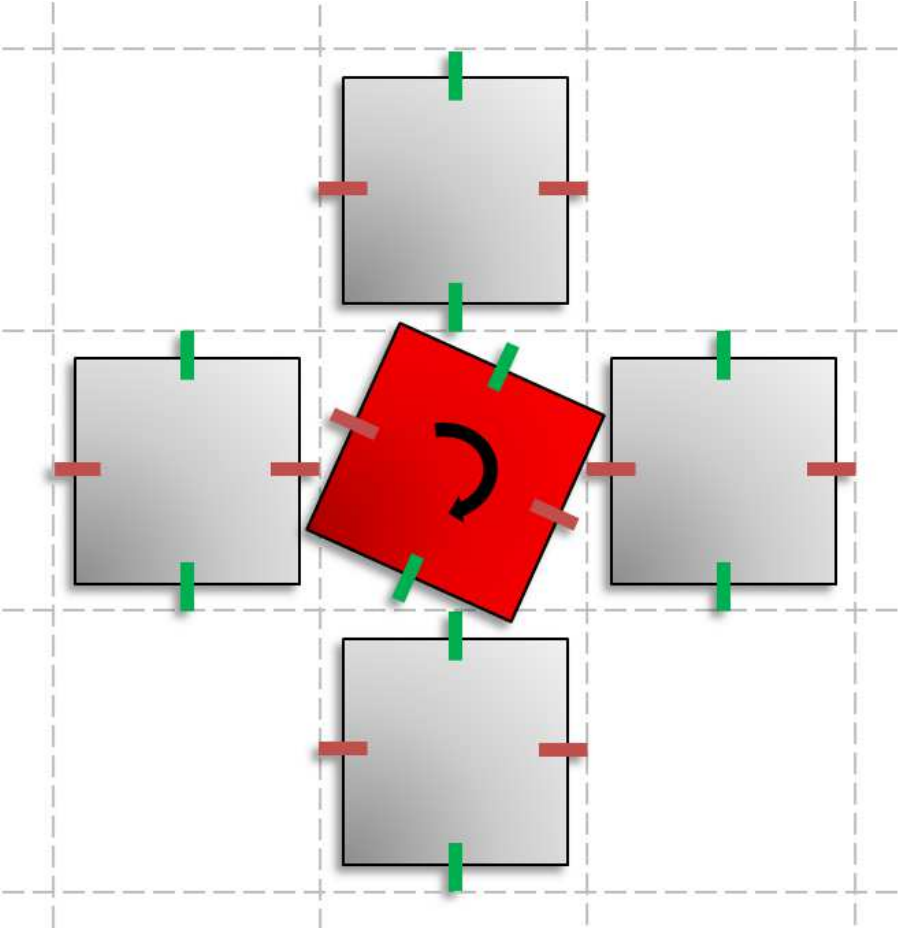}\\
(b) & (e)\\
[2ex]
\includegraphics[scale=0.4]{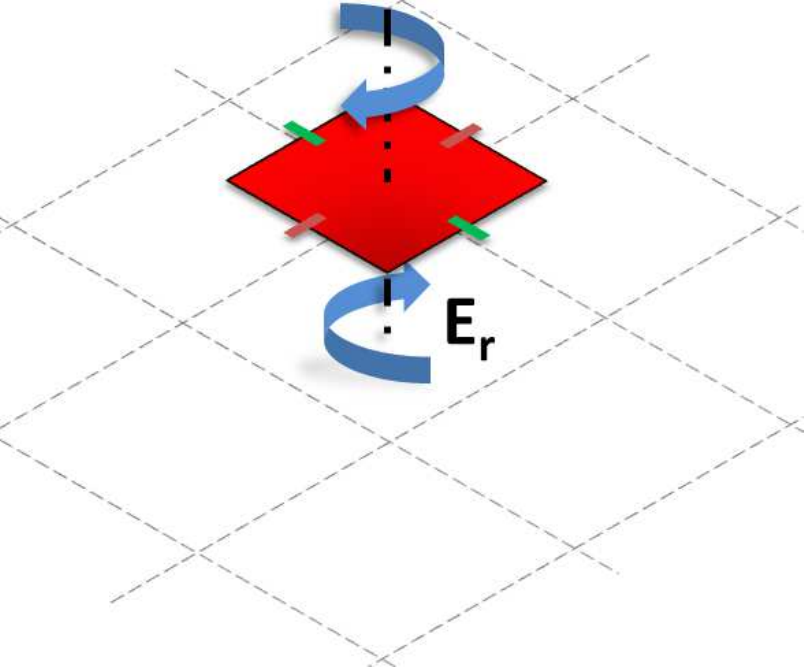} & \includegraphics[scale=0.3]{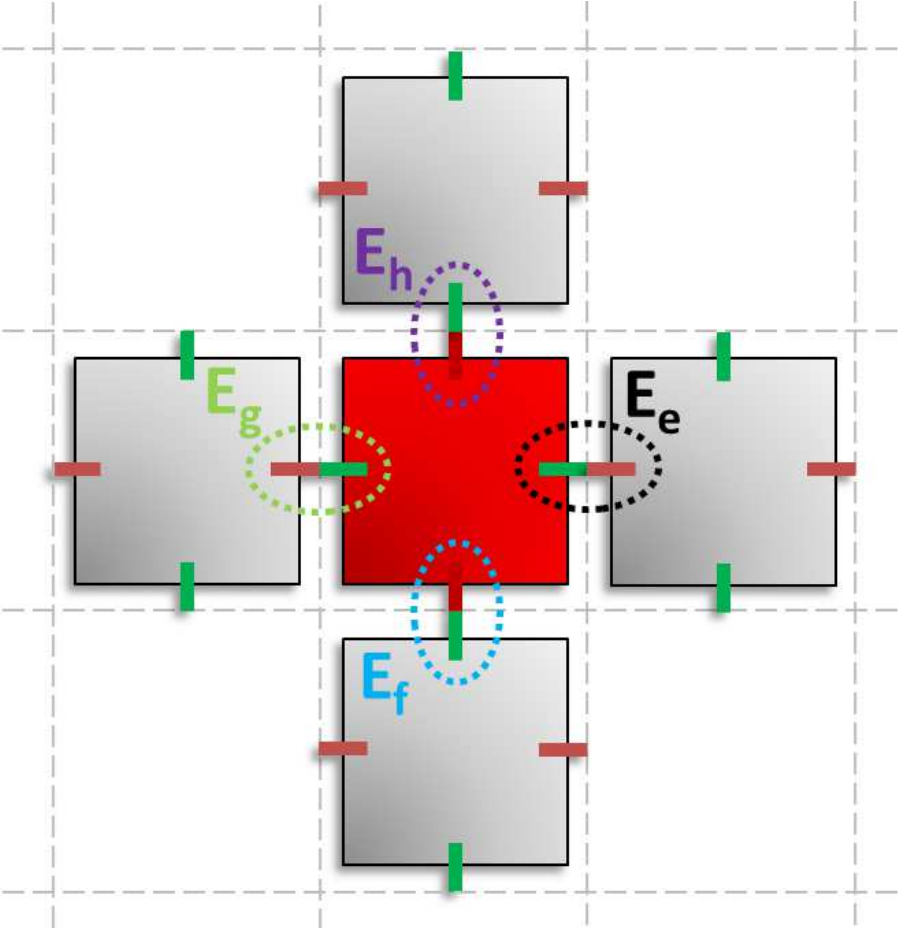}\\
(c) & (f)
\end{tabular}
\caption{\label{diffusion_rotation} \small Binding energy between a porphyrin molecule and substrate $E_s$ (a) and binding energies $E_a$, $E_b$, $E_c$ between a porphyrin molecule and its neighbours (b), all them involved when the molecule hops from position $(i,j)$ to $(i,j+1)$. The $+90$ degrees rotation of a porphyrin molecule involves binding energy of rotation $E_r$ (c), binding energy of starting configuration $E_a$, $E_b$, $E_c$, $E_d$ (d) and binding energies of final configuration $E_e$, $E_f$, $E_g$, $E_h$ (f). Differences in energy between starting and final configuration, i.e. $|E_a-E_f|$, $|E_b - E_g|$, $|E_c-E_g|$, $|E_d-E_e|$, are called energies of the saddle point.}
\end{figure}

The porphyrin-tiles kMC is configured with: 1) a set of porphyrin-tile families, or descriptors, each of these mapping a species of functionalised porphyrin molecule and from where porphyrin-tile instances are drawn to deposit on the lattice, and 2) numerical properties of the system. The latter comprises continuous numerical values to specify binding energies among functional groups, binding energy between a molecule and substrate, binding energy of rotation, concentration of each porphyrin-tile, environmental factors such as temperature of the system as well as discrete numerical values to specify number of labels and number of porphyrin-tile families. In each time step of the simulation, a list with the possible transitions of the system, i.e. deposition, motion and rotation of a porphyrin molecule, and their correspondent rates is compiled. In particular, depositions take place at a constant deposition rate ($R_{Dep}$) whereas diffusions and rotations are performed according to a diffusion rate ($r_{ijkl}$) calculated as: 
\begin{eqnarray}\label{diff_rate}
r_{ijkl} = \exp(\frac{-E_{ijkl}}{TT0})
\end{eqnarray}
where $E_{ijkl}$ is the activation energy a molecule needs to jump from position $(i,j)$ to position $(k,l)$ and $TT0$ is a fixed parameter capturing the temperature of the system and the Boltzmann constant. The activation energy for diffusion is calculated in terms of the binding energies involved between the porphyrin molecule of interest and each of its nearest neighbouring molecules and the binding energy to the substrate. For instance, the calculation of the activation energy for moving the porphyrin molecule in red colour of Fig. \ref{diffusion_rotation} (b) to the right is given by:
\begin{eqnarray}\label{eijkl_example_a}
E_{ijkl} = E_s + E_a c_1 + E_b c_2 + E_c c_3 
\end{eqnarray}
where $E_s$ is the binding energy between molecule and substrate (see Fig. \ref{diffusion_rotation} (a)), $E_{x\in \{a,b,c\}}$ is the binding energy between functional groups located at adjacent edges of neighbouring molecules and $c_{i \in \{1, 2, 3\}} \in \{0,1\}$ is the occupancy of neighbouring position $i$. Similarly, the activation energy for rotation is calculated in terms of the binding energies involved between the porphyrin molecule of interest and each of its nearest neighbours, the binding energy of rotation, and the binding energies of the saddle. For example, the calculation of the activation energy for rotating $+90$ degrees a porphyrin molecule like the one in red colour of Fig. \ref{diffusion_rotation} (d - f) is given by: 
\begin{eqnarray}\label{eijkl_example_b}
\begin{split}
E_{ijij} = E_r & + E_a c_1 \overline{c_2} + E_b c_2 \overline{c_3} + E_c c_3 \overline{c_4} + E_d c_4 \overline{c_1}\\
 & + | E_a - E_f| c_1 + | E_b - E_g| c_2\\
 & + | E_c - E_h| c_3 + | E_d - E_e| c_4
\end{split}
\end{eqnarray}
where $E_r$ is the binding energy for a porphyrin molecule to rotate $\pm 90$ degrees about its centre of mass (see Fig. \ref{diffusion_rotation} (c)) and $|E_x - E_y|$ is the energy of the saddle point. The latter is the difference between the binding energy of the breaking bond and the binding energy of the newly formed bonding. 

\begin{figure}[h]
\centering
\includegraphics[scale=1]{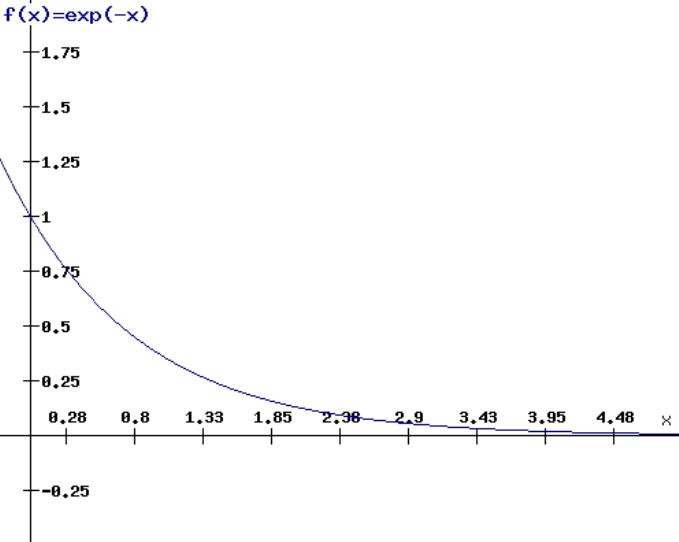}
\caption{\label{etothex} \small Plot of $r_{ijkl}$ defined in Eq. \ref{diff_rate} where $x$ is a scaled transformation of $E_{ijkl}/TT0$. During the Monte Carlo process, the bigger the activation energy, the lower chances its associated transition has to be chosen and performed.}
\end{figure}

Once the list of all possible transitions of the system and their rates are compiled, a Monte Carlo selection process follows in which the transition with the best chances to happen is chosen and performed. The chance of a transition is given according to the value of its associated rate which is directly linked to the activation energy. Hence, the bigger the activation energy of a transition, the lower chances it has to be performed. This rationale can be seen in the plot of Fig. \ref{etothex} where for explanation purposes $x$ is a scaled transformation of $E_{ijkl}/TT0$ in Eq. \ref{diff_rate}. Therefore, the more neighbouring molecules are present and the bigger $E_s$ or $E_r$, the lower chances a given porphyrin-tile has to diffuse or rotate. After a transition is performed, the list is updated and the process is repeated for a fixed number of time steps.

A pseudo-code of the porphyrin-tiles kMC system is listed in Algorithm \ref{kmc_alg}. The algorithm consists of three main data structures: $P$ that stores porphyrin-tiles, $T$ that lists the possible transitions of the system and the square site lattice $L$. The calculation of rates for motion and rotation associated to each porphyrin-tile takes place in {\small \textsc{calculateRates}} in which $activationEnergy(p, t)$ implements Eq. \ref{eijkl_example_a} when $t$ is motion or Eq. \ref{eijkl_example_b} when $t$ is rotation. The selection of the most likely transition takes place in {\small \textsc{doDiffusion}} where $performOn(t, p)$ moves or rotates a porphyrin-tile $p$. If neither rotation or motion takes place, the arrival of a new porphyrin-tile onto $L$ is performed by {\small \textsc{doDeposition}} where $getEmptyPosition(L)$ returns an empty site of the lattice.
\begin{algorithm}
\caption{porphyrin-tiles kMC algorithm}\label{kmc_alg}
\begin{algorithmic}[1]
\Procedure{Main}{$n$} 
\State $P \gets \varnothing$ 
\State $T \gets \{motion, rotation\}$
\For{$i\gets 1, n$}
  \State $calcRates$
  \If {$\mathbf{not}$($doDiffusion$)}
    \State $doDeposition$
  \EndIf
 \EndFor
\EndProcedure
\\
\Procedure{calculateRates}{}
\ForAll{$p \in P$}
  \ForAll{$t \in T$}
    \State $E_{ijkl} \gets activationEnergy(p,t)$
    \State $p.r_t \gets exp(-E_{ijkl}/TT0)$
  \EndFor
\EndFor
\EndProcedure
\\
\Procedure{doDiffusion}{} 
\ForAll{$p \in P$}
  \State $total \gets total + p.r_t$
\EndFor
\State $counter \gets 0$
\State $rndShoot \gets RND[0,1) \times (total + r_{Dep})$
\ForAll{$p \in P$}
  \State $counter \gets counter + p.r_t$
  \If{$counter > rndShoot$} 
	\State $performOn(t, p)$
	\State \Return $TRUE$
  \EndIf
\EndFor
\State \Return $FALSE$
\EndProcedure
\\
\Procedure{doDeposition}{} 
\State $p_{new}.ij \gets getEmptyPosition(L)$
\State $P \cup \{p_{new}\}$
\EndProcedure
\end{algorithmic}
\end{algorithm}

In what follows we present the experiments and results obtained with the porphyrin-tiles kMC system. The aim here is to explore what self-assembled aggregates are possible to obtain as we vary the binding energy between porphyrin molecule and substrate, and as we program porphyrin-tiles with different functional groups. 

\subsection{Experiments} \label{pt_kmc}
For each experiment, the simulator was configured with: a lattice of $256 \times 256$ positions, two different species of heterogeneous iso-functionalised porphyrin-tiles, binding energy between each of the two identical functional groups ($E_{11}$ and $E_{22}$), binding energy between different functional groups ($E_{12}$) and binding energy between molecule and substrate ($E_s$). The first three binding energies taking values from $[0.1\,eV, 0.2\,eV, \dots, 1.0\,eV]$ whilst the latter from $[0.5\,eV, 0.6\,eV, \dots, 1.0\,eV]$. Although the programmability of porphyrin-tiles is conceptually given by changing the functional groups assigned to the labels representing structural units, its actual implementation is carried out by changing the values assigned to $E_{11}$, $E_{22}$ and $E_{12}$. Thus, all the possible combinations among $E_{11}$, $E_{22}$, $E_{12}$ and $E_s$ were systematically given in turns together with maximum lattice coverage of $25\%$, $E_r=1.3\,eV$, $TT0=28\times10^{-3}$ and $R_{Dep}=5\times10^{-5}$. From now onwards the units of energy $\,eV$ will be omitted when referring to values taken by $E_s$, $E_{11}$, $E_{22}$ and $E_{12}$. 

\subsection{Results} \label{sec_results}
\begin{figure}[h]
\centering
\includegraphics[scale=0.9]{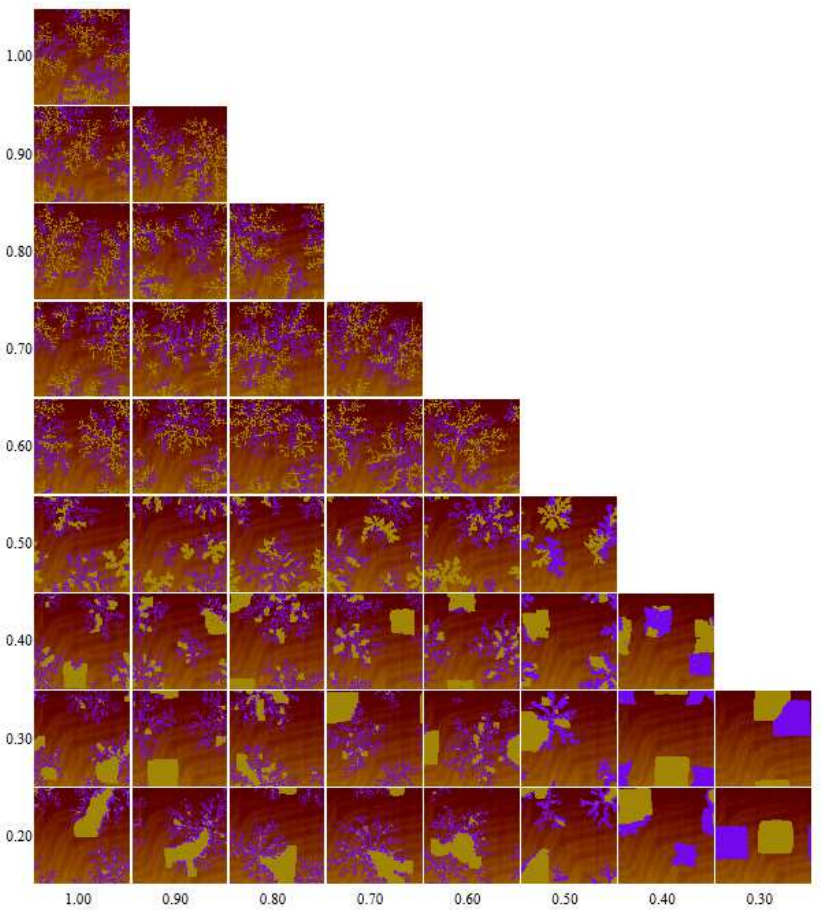}
\caption{\label{results_sample} \small A sample of experiment results set with two species of heterogeneous iso-functionalised porphyrin nano-tiles (yellow and blue), $E_{11} \in [0.3\,eV, 0.4\,eV, \dots, 1.0\,eV]$, $E_{22} \in [0.2\,eV, 0.3\,eV, \dots, 1.0\,eV]$, $E_{12}=0.1\,eV$ and $E_s=0.5\,eV$.}
\end{figure}

The final configuration of each experiment was captured in an image available for inspection in a website at \url{http://www.cs.nott.ac.uk/~gzt/3x1St3nC3}. We observe that the four input parameters have different levels of impact on the simulation results. To begin with, the binding energy between molecule and substrate controls the quantity of originated aggregates. That is, the smaller (bigger) the value of $E_s$, the lower (higher) stickiness to the substrate that gives a molecule more (less) freedom to move. Second, we observe that by programming the porphyrin-tiles in specific ways it is possible to control the aggregates' composition, morphology as well as diversity. For instance, the binding energy between different functional groups influences the composition of the aggregates and diversity of morphologies. In other words, more (less) segregation between porphyrin-tiles species as well as more (less) diversity on aggregates morphology is observed when $E_{12}$ increases (decreases). In addition, the combinations of binding energies between identical functional groups has a direct impact on the morphology of the aggregates. These binding energies are, however, more (less) influential in the presence of low (high) $E_s$ and low $E_{12}$. A representative selection of simulation results is shown in Fig. \ref{results_sample} where the number of aggregates in each simulation result goes between $1$ and $5$ due to low $E_s$. A visual inspection reveals that there is more segregation in the aggregates composition since low $E_{12}$ drives the self-assembly process towards the creation of aggregates among molecules of the same species. From a more general point of view, there is also a transition on the morphologies of the aggregates which goes from square bulky (Fig. \ref{results_sample} bottom right) to thin dendritic (Fig. \ref{results_sample} top left) as $E_{11}$ and $E_{22}$ vary across their range.

\section{Kolmogorov Complexity and Information Content}

Algorithmic complexity \citep{Kolmogorov1965QuantitativeInformation,Chaitin69onthe} characterises the information content of an object as the shortest computer program that produces it. The result of the difference between the length of a string and its greatest compressed version determines the complexity of a string and how difficult it is to predict. Formally,
\begin{equation}
K_U(s) = min\{|p|, U(p)=s\}
\end{equation}
where $|p|$ is the length of the shortest program that produces a string $s$ running on a universal Turing machine $U$. The minimal length for a description of an object depends on the exact method used for reproducing the object from the description, but the Invariance theorem guarantees that differences will be bounded by a constant, coincide in the limit and do not depend on the object. More formally, the theorem establishes that if $U_1$ and $U_2$ are two (universal) Turing machines and $K_{U_1}(s)$ and $K_{U_2}(s)$ algorithmic complexities of a binary string $s$ when $U_1$ or $U_2$ are respectively used, there exists a constant $c$ such that for all binary strings $s$:
\begin{equation}
| K_{U_1}(s) - K_{U_2}(s) | < c_{_{U_1,U_2}}
\end{equation}

No algorithm can tell whether a program $p$ generating $s$ is the shortest (due to the undecidability of the halting problem of Turing machines) but $K(s)$ is upper semi-computable meaning that it can be approximated from above, for example, using lossless compression algorithms. The result of a compression algorithm is a sufficient test for non-randomness, i.e. $K$ cannot be greater than the length of the compressed version ($C$) of $s$. Previous investigations on a phase transition coefficient \citep{zenil10compressionbased,Zenil12onthe} were undertaken in an attempt to quantify the qualitative behaviour of systems with order parameters based on these notions, namely Kolmogorov complexity and compressibility. The implementations of such concepts have been performed in terms of the Deflate compression algorithm which is available in both \emph{pngcrush} (zlib) tool\footnote{Available at \url{http://pmt.sourceforge.net/pngcrush/} (Accessed on October 14, 2012) set to maximum compression.} version and the function \emph{Compress} in \emph{Mathematica} v.8. Deflate is a variation of the universal \citep{Li2004} lossless data compression algorithm LZ77 (Lempel-Ziv) popular in many computer formats such as Portable Network Graphics (PNG) and GNU zip (gzip).

In what follows, Kolmogorov complexity notions are applied to the simulation results obtained from experiments performed with the porphyrin-tiles kMC system presented in Section \ref{pt_kmc}. Our interest in employing these concepts here is two-fold. First, we would like to investigate if there exists any type of correlation between the input parameters values of the system and the simulation results. Also, we are interested to investigate if phase transitions emerge across the complexity associated to the captured structures. The second goal is to apply a similarity measure based on the notion of information distance in order to automatically classify the self-assembled aggregates according to their information content.
\begin{figure*}[ht!]
\centering
\scalebox{0.47}{\includegraphics{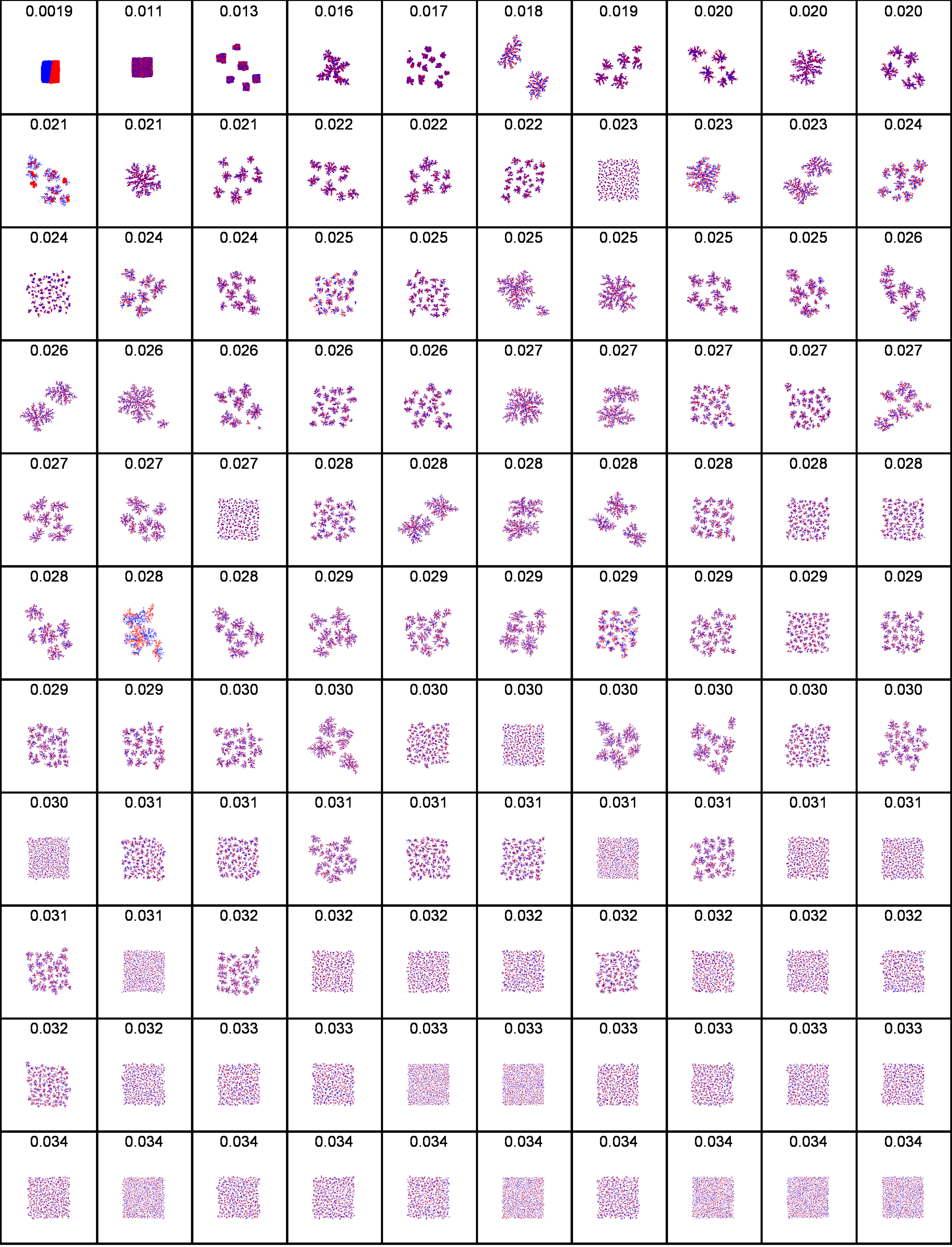}}
\caption{\label{comp_classification}\small Incompressibility classification: Aggregate configurations of porphyrin-tiles sorted by compression ratio (from greatest to lowest compression). It is clear that aggregates with similar qualitative structural properties are close to each other, meaning that the compressibility measure is capturing behavioural traits of the configurations in which the porphyrin-tiles distribute themselves when interacting under different parameter conditions.}
\end{figure*}

\subsection{Compression-based analysis} \label{comp_analysis}

A particular combination of binding energy values, i.e. combinations of $E_s$, $E_{11}$, $E_{22}$ and $E_{12}$, define a point in the input parameter space which, after the phorphyrin-tiles kMC simulation finishes, links to another point onto the simulation result space. Here, we define this last space in terms of compressibility which is a measure defining how compressible or incompressible a given input is. In general, a string is called {\it compressible} if it has a description which is much shorter than the string itself. Conversely, an {\it incompressible} string lacks regularities that could be exploited to obtain a compressed description; they are patternless, hence random. Thus, a simulation result is captured in an image which, seen as a collection of strings arranged in a special way, works as input to a compression method. The output of such method is a compressed file associated to a {\it compression ratio} defined as the compression size of the image divided by the size of its uncompressed form. In particular, the more compressible the image is, the smaller the compression ratio of the associated compressed file. Having all the resulting experiments captured in images we would like to address the following:\\\\
{\it Is it possible to classify the simulation results in terms of compressibility ? If so, could phase transitions be discovered from such classification ?\\\\ Is there any correlation between input parameter space and the simulation result space ?}\\

In order to answer the first question, we employed compressibility analysis. That is, each image capturing a simulation result is compressed with the Deflate algorithm implemented in pngcrush. This returns as a result a compressed PNG file the size of which reveals its compressibility. In other words, the smaller (bigger) the compression ratio of the PNG file, the more compressible (incompressible) the captured simulation is. The collection of images shown in Fig. \ref{comp_classification} are some representatives of the experiments. From left to right and top to bottom, these images are sorted in ascending order according to their associated compression ratio. These findings reveal that simulation results with similar qualitative structural properties among their aggregates are close to each other, meaning that the compressibility measure is capturing behavioural traits of the configurations in which the porphyrin-tiles distribute themselves when interacting under different input parameter conditions.
\begin{figure}[h]
\centering
\scalebox{.30}{\includegraphics{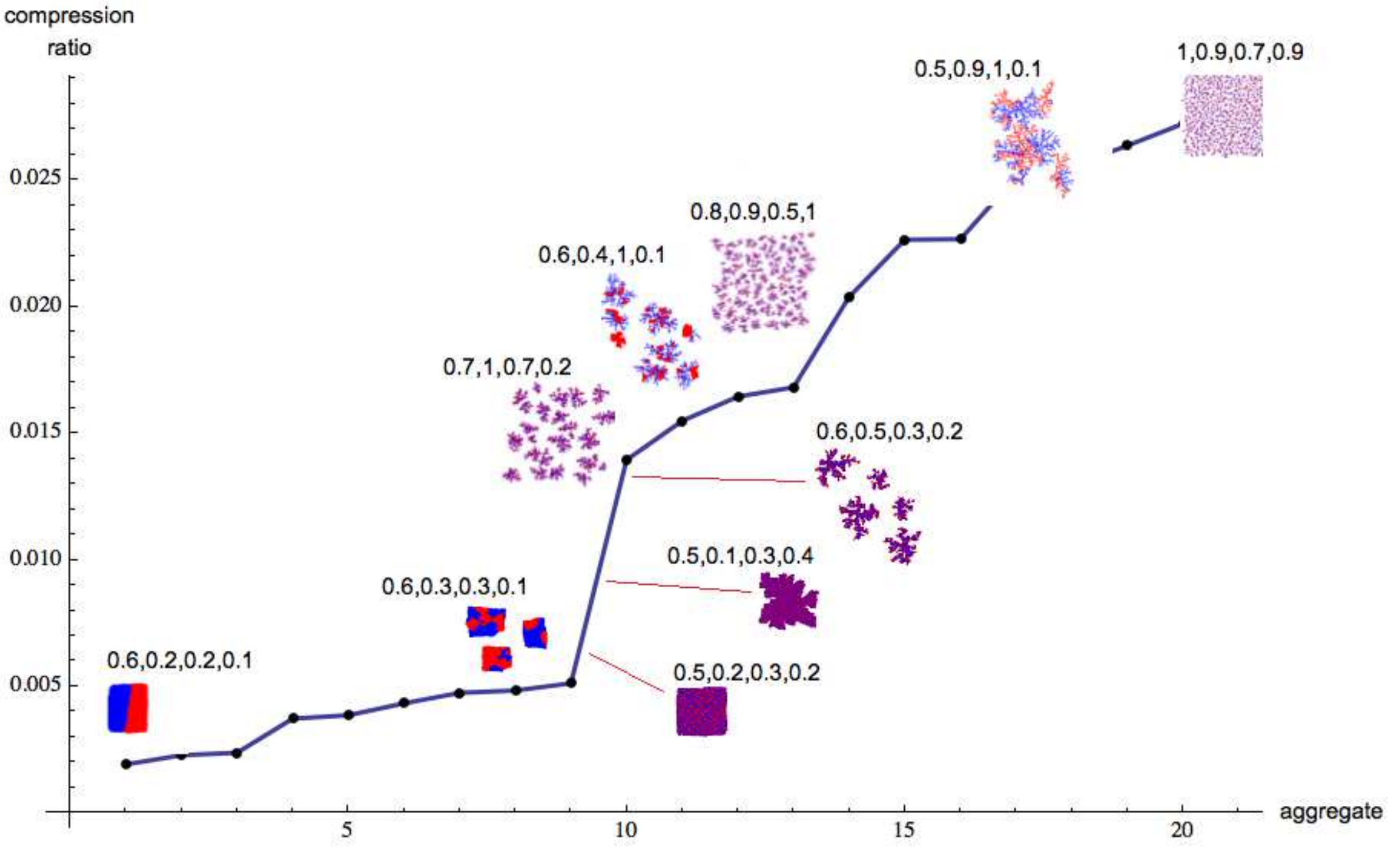}}
\caption{\label{comp_phase_transition} \small As a proof-of-concept in the application of Kolmogorov complexity to the conformation space of Porphyrin molecules, a sample of $21$ representative aggregates with qualitative different behaviour is selected with the aid of an agglomerative hierarchical clustering with an Euclidian distance function. A phase transition appeared between highly compressible configurations (aggregates $1$ to $9$) and low compressible configurations (aggregates $10$ to $21$) that turned out to be a transition of fast changes according to the compression ratios, as shown in Fig. \ref{comp_distribution}. Aggregates are labelled with their respective $E_s$, $E_{11}$, $E_{22}$, $E_{12}$.}
\end{figure}

Considering the entire set of simulation results, we are interested to see if it was possible to distinguish phase transitions in terms of compressibility. Thus, we ran a hierarchical clustering algorithm \citep{Anderberg-Cluster-1973} (see further details in Fig. \ref{comp_phase_transition}) leading to the $21$ groups from which a (uniformly) random element was chosen. A steep rise was observed revealing a sharp increment in the complexity of the self-assembled aggregates. As explained in Section \ref{sec_results}, input parameter values act as ``microscopic'' interactions which give origin to ``macroscopic'' manifestations. A richer analysis on how these relates to compressibility is done in Section \ref{param_disc_ortho}.
\begin{figure}[h]
\centering
\scalebox{.27}{\includegraphics{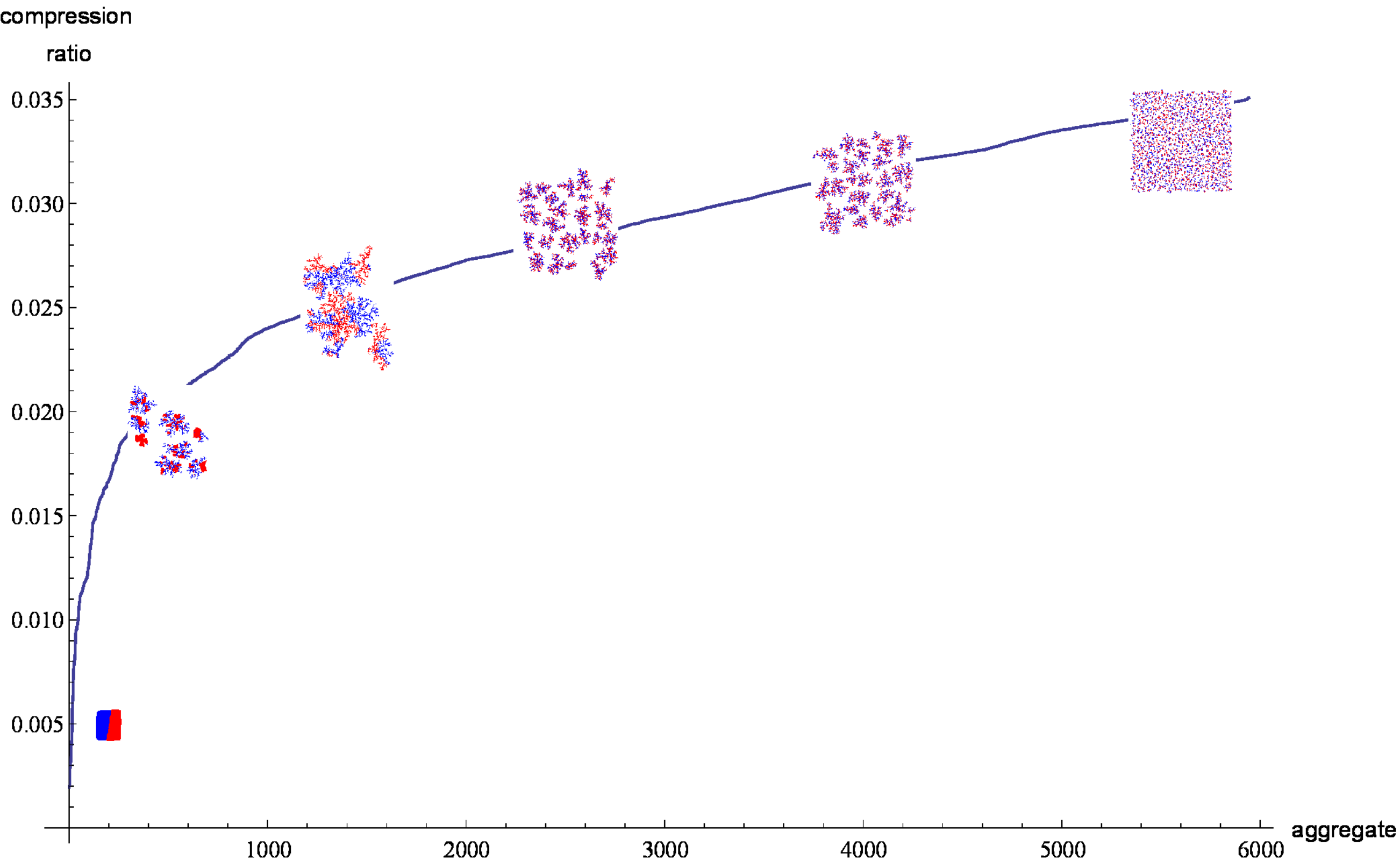}}
\caption{\label{comp_distribution}\small Looking at the distribution of compression ratios one can see that aggregate configurations quickly reach medium complexity before evolving into more random configurations. The distribution of aggregates is also conform with an intuition of increasing randomness (Kolmogorov complexity), where simple (complex) configurations are close (far) to the origin.}
\end{figure}

We next sort the aggregates in terms of compression ratio as shown in Fig. \ref{comp_distribution}. The resulting plot reveals that there are two relevant segments within compression ratios, one going from $0.0019$ to $0.020$ and another one going from $0.020$ to $0.034$. Within the first one, there is a fast evolution going from low to medium complexity structures whereas within the second segment there is a rather smooth transition from medium towards high complexity structures. In addition, an increase in randomness along x-axis is observed along the entire distribution of self-assembled aggregates where the simpler the structures, the closer to the origin.

In order to address the second question, the notion of distance within the input parameter space is defined as follows. Let a point in the input parameter space be defined as a 4-tuple comprising the binding energies associated to a simulation result, i.e. $P_i=(E_s, E_{11}, E_{22}, E_{12})$. Let the point where all binding energies have zero value the origin of the input parameter space, i.e. $P_o = (0, 0, 0, 0)$. The distance of a point $P_i$ in the input parameter space is defined as the Euclidean distance between $P_o$ and $P_i$, formally speaking:
\begin{equation}\label{euc_dist}
dist(P_i) = \sqrt{E_s^2 + E_{11}^2 + E_{22}^2 + E_{12}^2}
\end{equation}

\begin{figure*}[ht!]
\centering
\scalebox{0.35}{\includegraphics{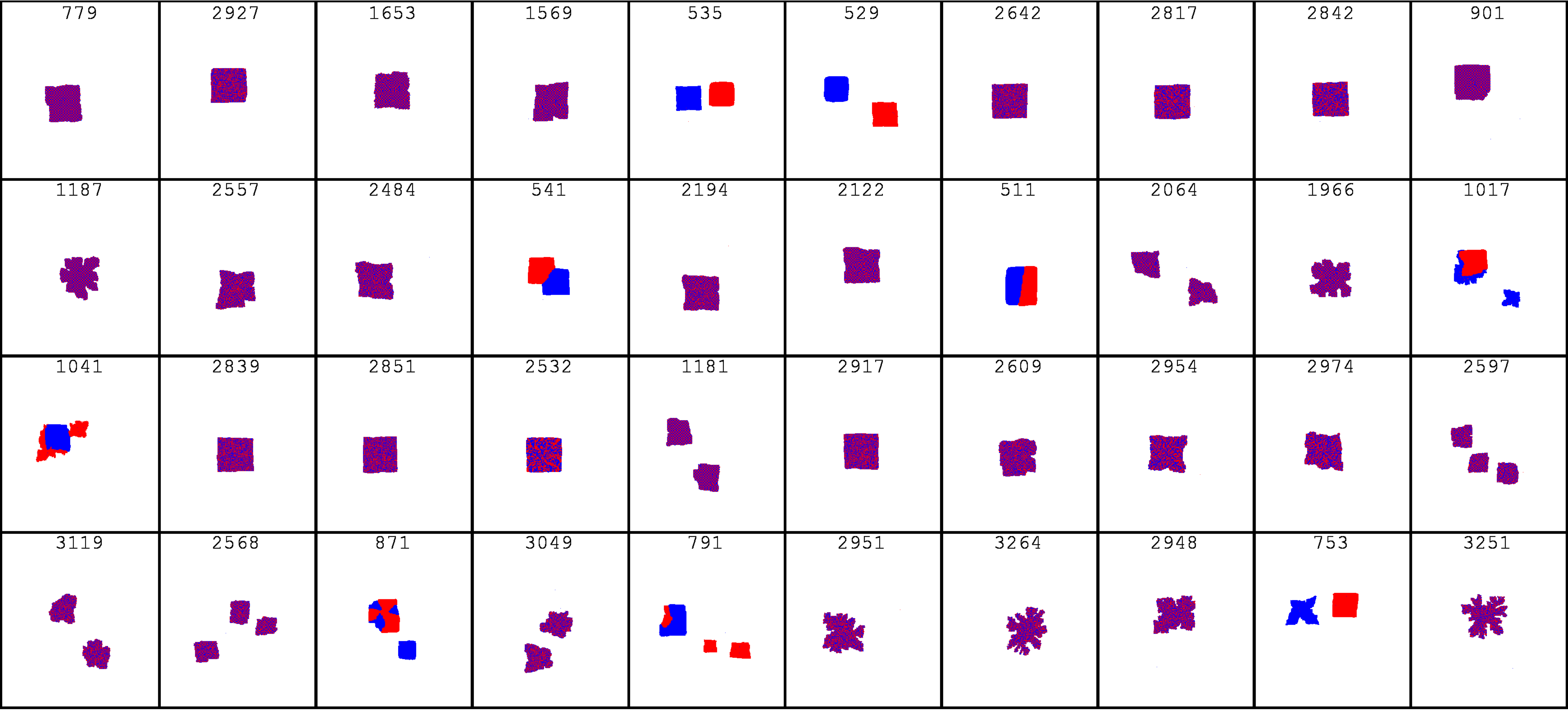}}
\caption{\label{close_to_po_sample}\small Aggregates close to the parameter-space origin (where all parameters have zero values) are highly compressible (low Kolmogorov complexity). Images are preceded by their compressibility value (in bits).}
\end{figure*}

For each simulation result, its associated distance to the origin within the input parameter space and its approximate Kolmogorov complexity were calculated. The latter as an approximation by measuring the size of the compressed file when applying Deflate algorithm implemented in pngcrush. After sorting the simulation results by distance to the origin within input parameter space in ascending order, it was reveled that there exists a correlation between the input parameter space and the simulation result space. In particular, we observe that the farther (closer) to the origin a simulation result, the higher (lower) its associated compression size. As an example, Fig. \ref{close_to_po_sample} shows from top to bottom and left to right, the first $40$ simulation results located close to the origin, sorted by distance within input parameter space in ascending order and labelled with their estimated Kolmogorov complexity. 

\subsection{Information content analysis}

Although a general correlation between input parameter space and simulation result space has been defined in terms of compression ratio, nothing is yet said about distance between the constituents of the simulation result space. The \emph{Normalized Information Distance} based on the Kolmogorov complexity was proposed in \cite{Li2004} and shown to be \emph{universal} in the sense that it discovers all computable similarities. A computable version was also suggested in \cite{Li2004}, called \emph{normalized compressed distance} ($NCD$) defined as:
\begin{equation}
NCD(o_1,o_2)=\frac{C(o_1o_2)-\min\{C(o_1),C(o_2)\}}{\max\{C(o_1),C(o_2)\}}
\end{equation}
where $C$ is the computable compressed length (in bits) of input $o_i$, and $o_1o_2$ is the concatenation of $o_1$ and $o_2$. Successful applications of $NCD$, such as classification of mtDNA sequences, can be found in \cite{Li2004}. To the knowledge of these authors, we make use of these measures to questions of synthetic biology in this paper for the first time. Our aim here is to employ $NCD$ in order to study the distance in terms of information content embedded in the self-assembled aggregates captured by the simulation results, in particular we seek to address the following question:\\\\
{\it Is it possible to characterise the simulation results and the distribution of their self-assembled aggregates in terms of information content ?}\\

\begin{figure}[t!]
\centering
\scalebox{.20}{\includegraphics{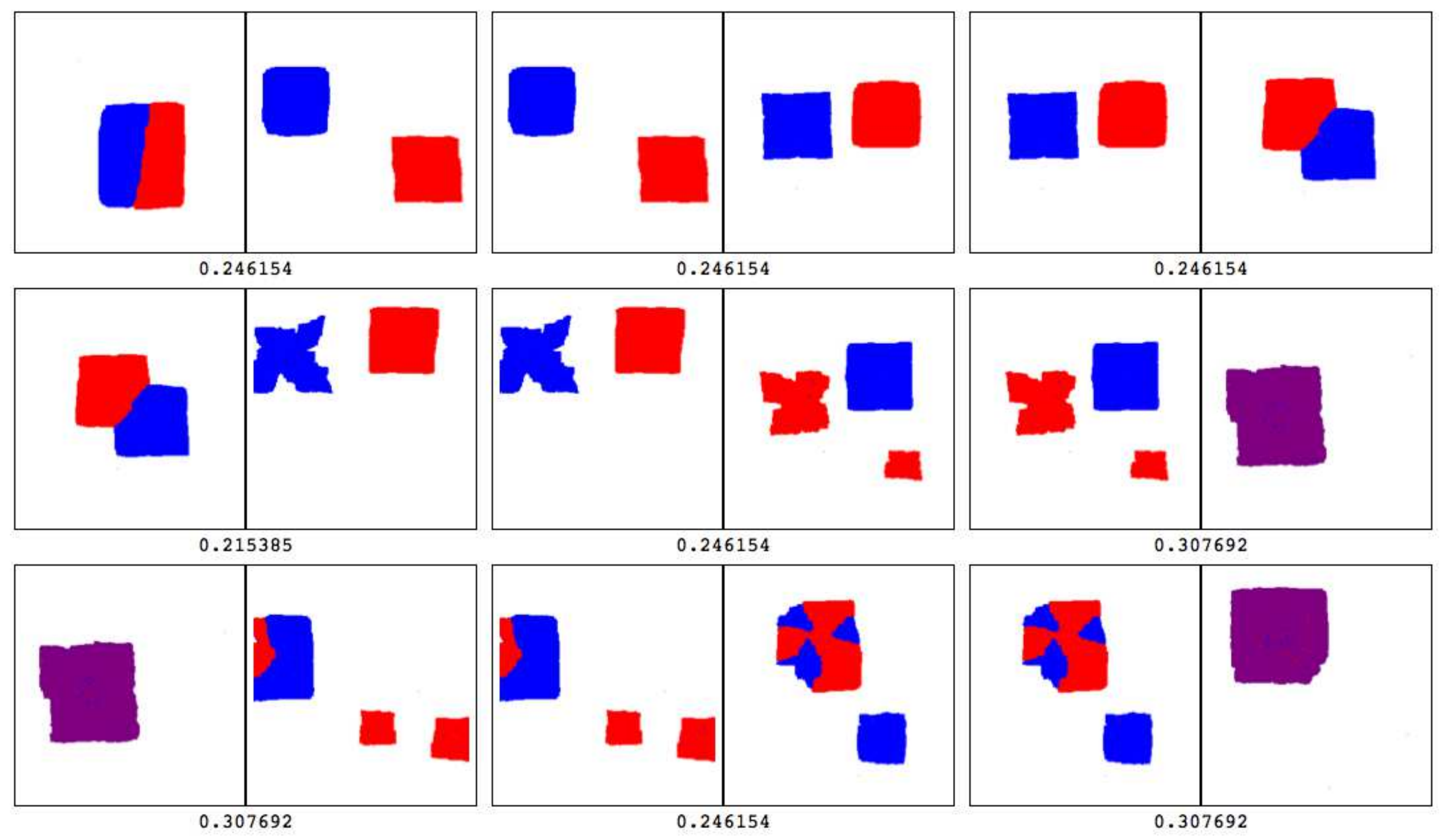}}
\caption{\label{ncd_close_to_origin}\small Aggregates close to each other in the parameter-space origin are compressible (low Kolmogorov complexity) and NCD is less sensitive to capture similarity.}
\end{figure}

\begin{figure}[h!]
\centering
\scalebox{.20}{\includegraphics{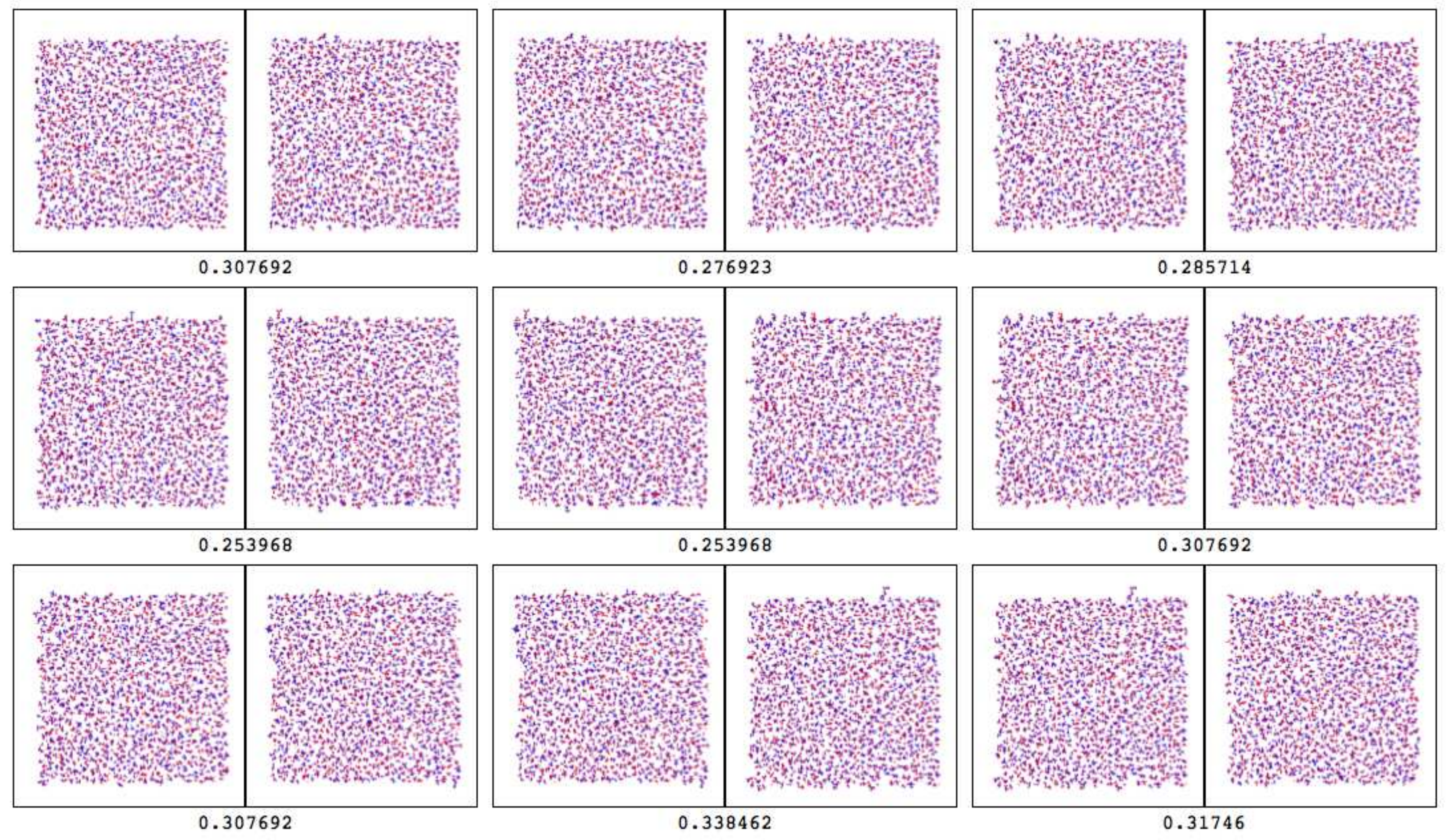}}
\caption{\label{ncd_far_to_origin}\small Aggregates far from the parameter-space origin are poorly compressible (high Kolmogorov complexity) but also have small $NCD$ values (high similarity).}
\end{figure}

To begin with, we apply $NCD$ among samples of simulation results located close to the origin of the input parameter space and also among those which are far from the origin of the input parameter space. The idea is to see how $NCD$ works when applied among highly compressible simulation results and among low compressible ones. For the first experiment, we arrange the simulation results shown in Fig. \ref{close_to_po_sample} according to their compression size in ascending order and from these the first ten are taken. Then, all the pairwise combinations between consecutive simulation results were set as input to $NCD$ the output of which is depicted in Fig. \ref{ncd_close_to_origin}. The findings reveal that highly compressible simulations which are close to each other in the input parameter space receive small $NCD$ values, hence indicating that they are similar to one another and share information content. Similarly, the experiment conducted among low compressible samples also reveals that close to each other simulation results receive small $NCD$ values as depicted in Fig. \ref{ncd_far_to_origin}.

\begin{figure}[h]
\centering
\scalebox{.25}{\includegraphics{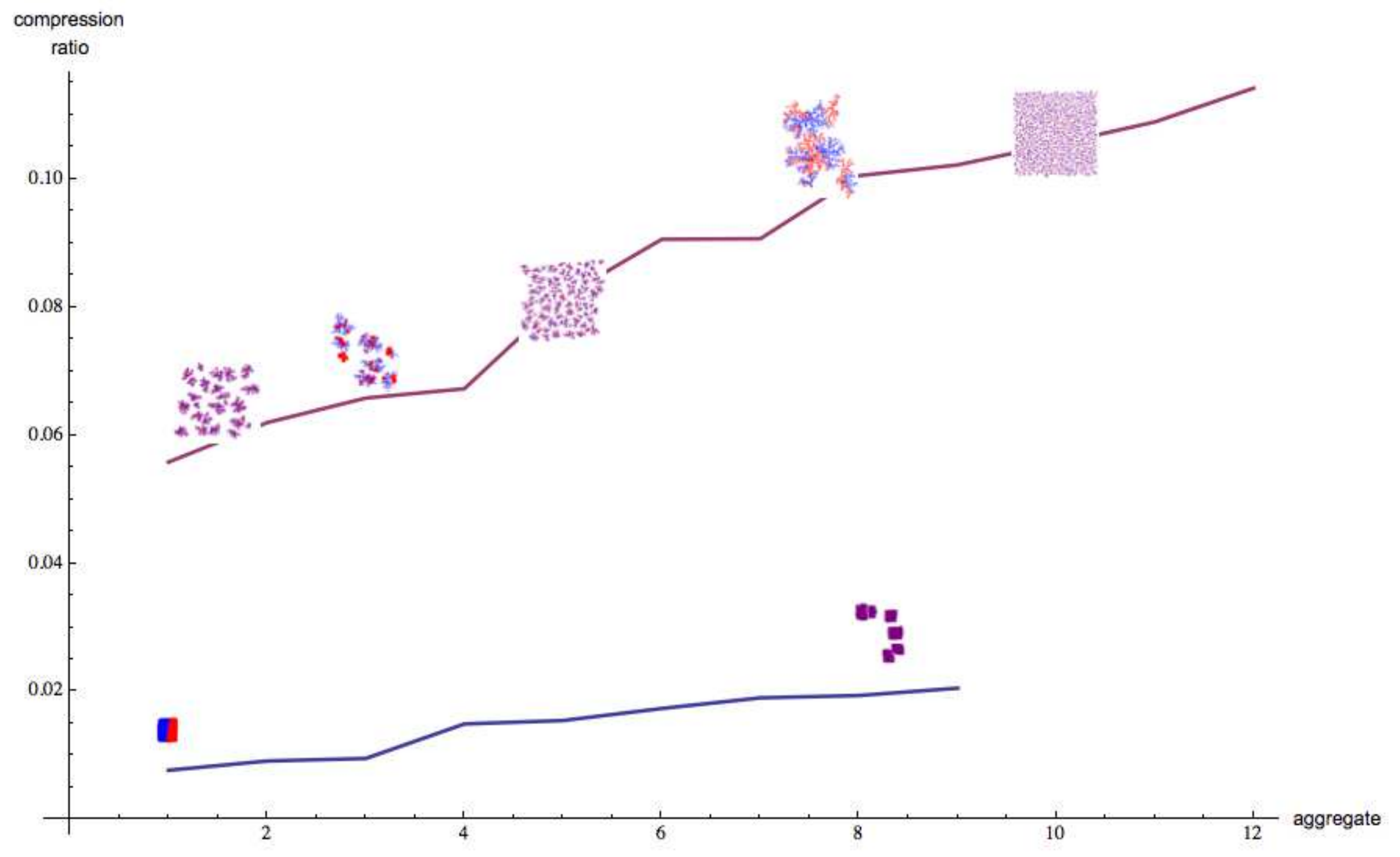}}
\caption{\label{ncd_clustering}\small A simple clustering algorithm using $NCD$ as distance function yields 2 clear groups formed by aggregates with high compression ratio (low Kolmogorov complexity) and low compression ratio (higher Kolmogorov complexity).}
\end{figure}

Considering the entire set of simulation results, we are now interested to see if it is possible to distinguish groups among the simulation results by means of information content. In order to do this, the $21$ representatives of Section \ref{comp_analysis} are employed to build a distance matrix using $NCD$ as distance function. This distance matrix is then set as input for a simple clustering algorithm. The clustering reveals two different groups, one with high compression ratio and another one with low compression ratio. These groups as well as their constituent samples sorted by compression ratio are depicted in Fig. \ref{ncd_clustering}.

\subsection{Variability}

\begin{figure}[h] 
\centering
\scalebox{0.30}{\includegraphics{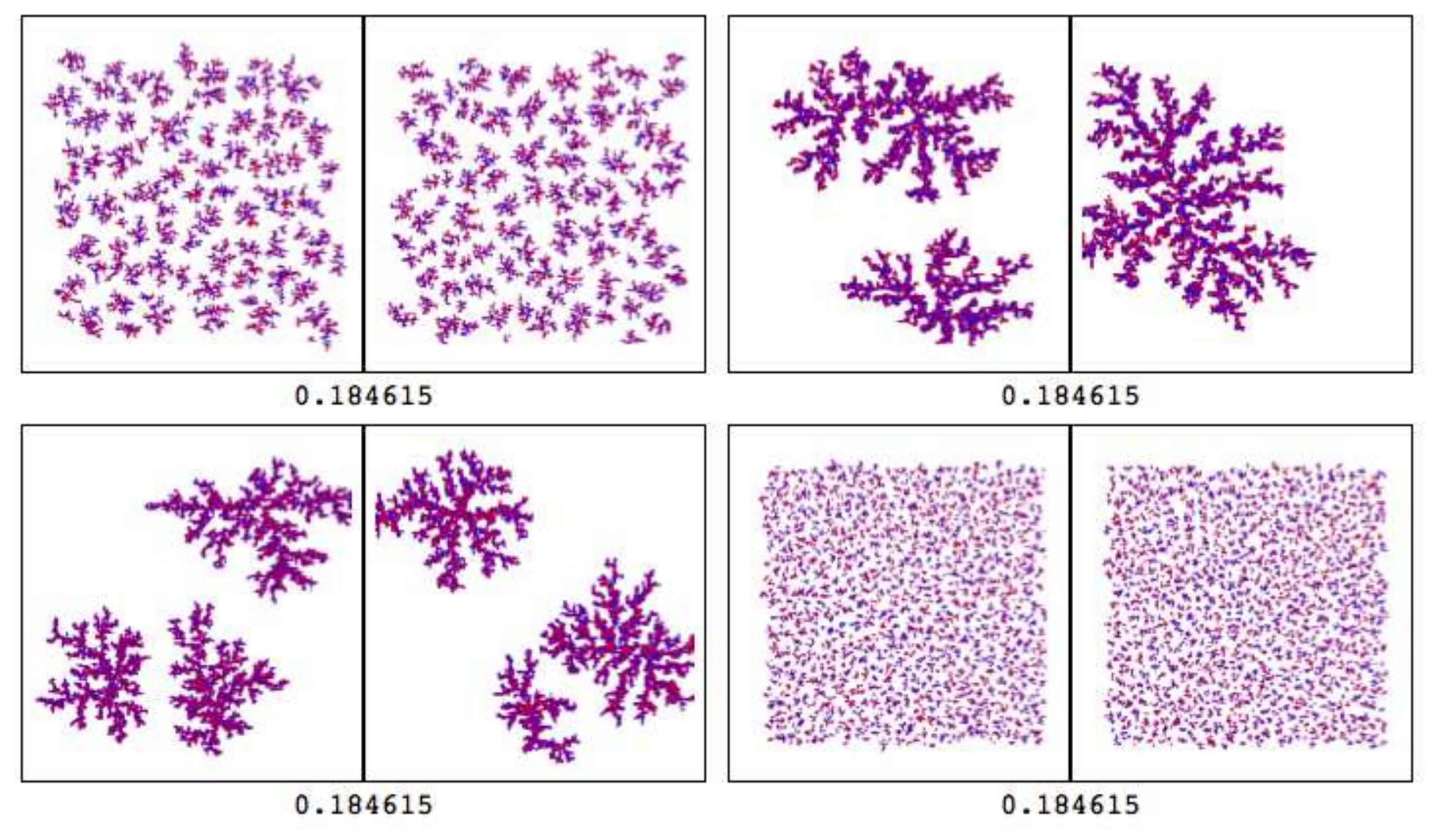}}
\caption{\label{ncd_small}\small Consecutive aggregates by compression ratio with small $NCD$ show that compression ratio and $NCD$ are both characterising similarity.}
\end{figure}

\begin{figure}[h]
\centering
\scalebox{0.20}{\includegraphics{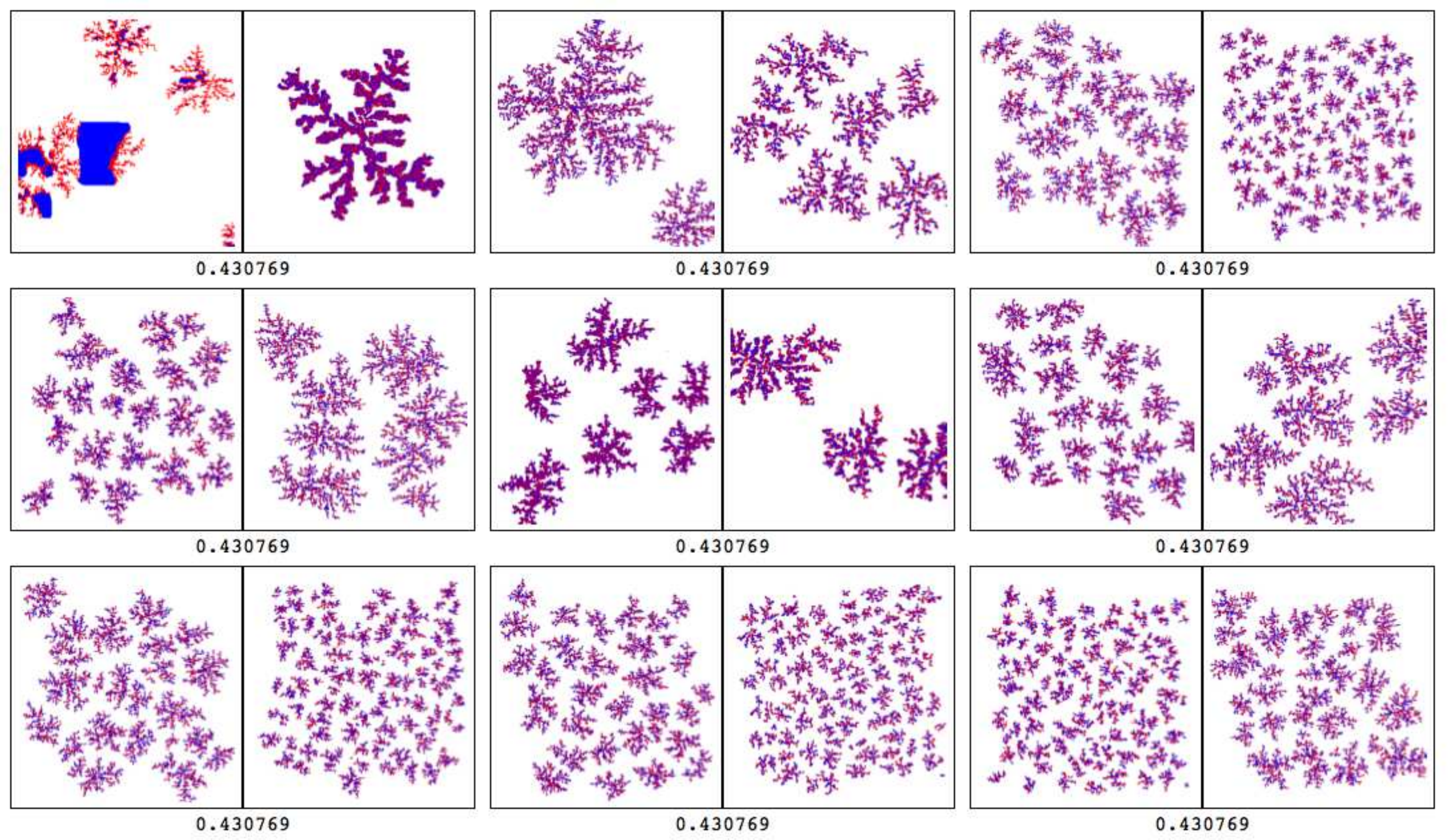}}
\caption{\label{ncd_big}\small Consecutive aggregates by compression ratio that have greatest $NCD$ provide another tool to detect possible phase transitions.}
\end{figure}

\begin{figure}[hb!]
\centering
\scalebox{0.22}{\includegraphics{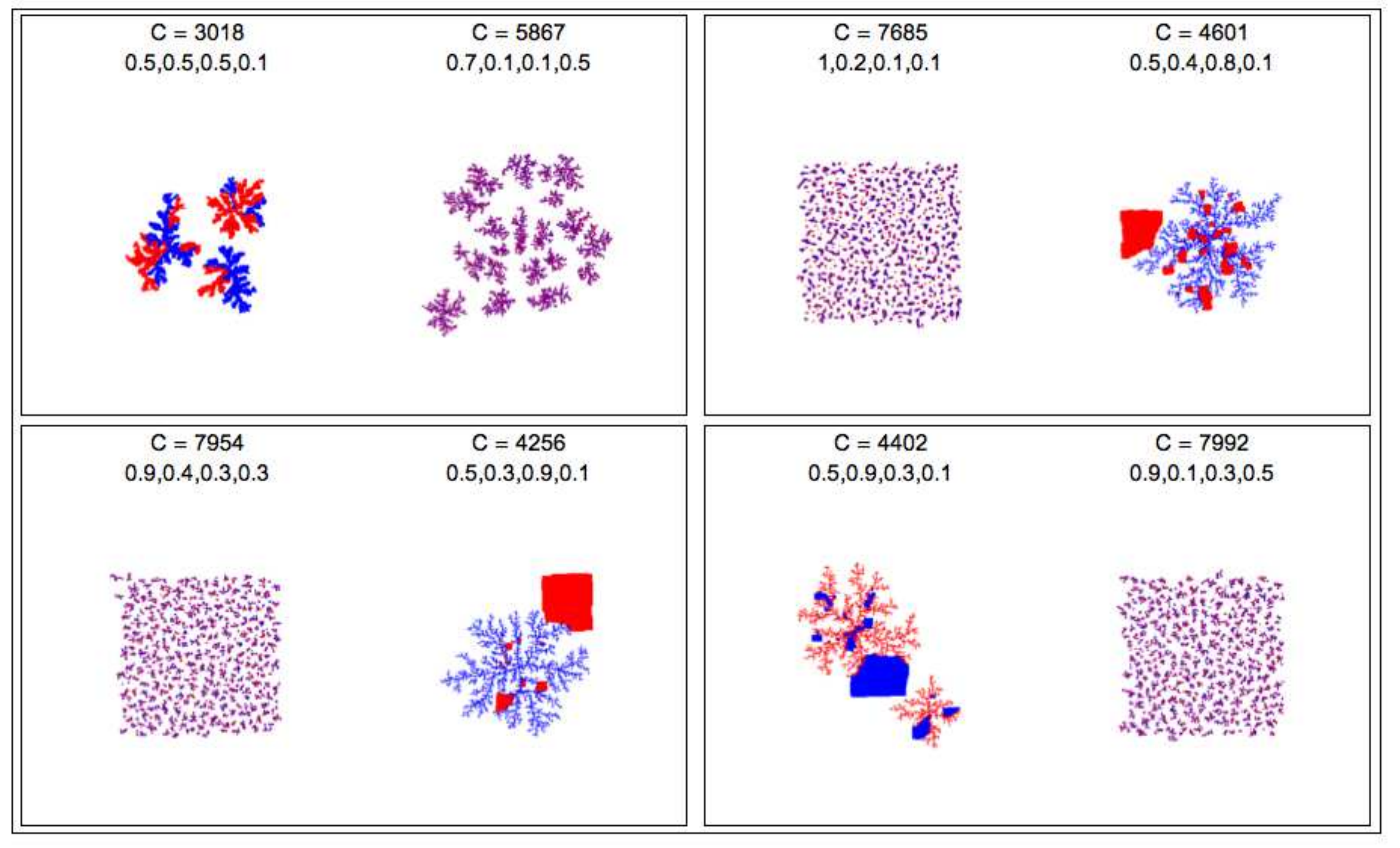}}
\caption{\label{comp_small}\small Aggregates sorted by Euclidean distance to the origin (where all parameters equal to zero) that have largest compression length differences ($|C(o_1) - C(o_2)| > 2500$ bits).}
\end{figure}

Compression-based as well as information content analysis applied to the simulation result space allowed us to characterize both behaviour and distribution of the captured self-assembled aggregates. Examples of such are given in Figs. \ref{ncd_small} and  \ref{ncd_big} which show cases of consecutive aggregates sorted by compression ratio presenting the smallest and largest $NCD$ respectively. In fact, the results observed here suggest that $NCD$ could be employed as an alternative route to discover phase transitions in terms of information content among the entire set of simulation results. Likewise, equivalent examples can be found when simulation results are sorted in terms of the Euclidean distance defined in Eq. \ref{euc_dist}. For instance, Fig. \ref{comp_small} shows pairs of consecutive simulation results which have the largest compression length differences. One of the aims would then be to find a suitable algorithmic measure for every qualitative trait of a system capturing the behaviour of it. Clearly, the idea of ultimately programming a system is related to the variability of a system given that a system with no apparent variability cannot be programmed. Programmability is hence both a combination of behavioural change and external control. 

\begin{figure*}[ht]
\centering
\scalebox{0.35}{\includegraphics{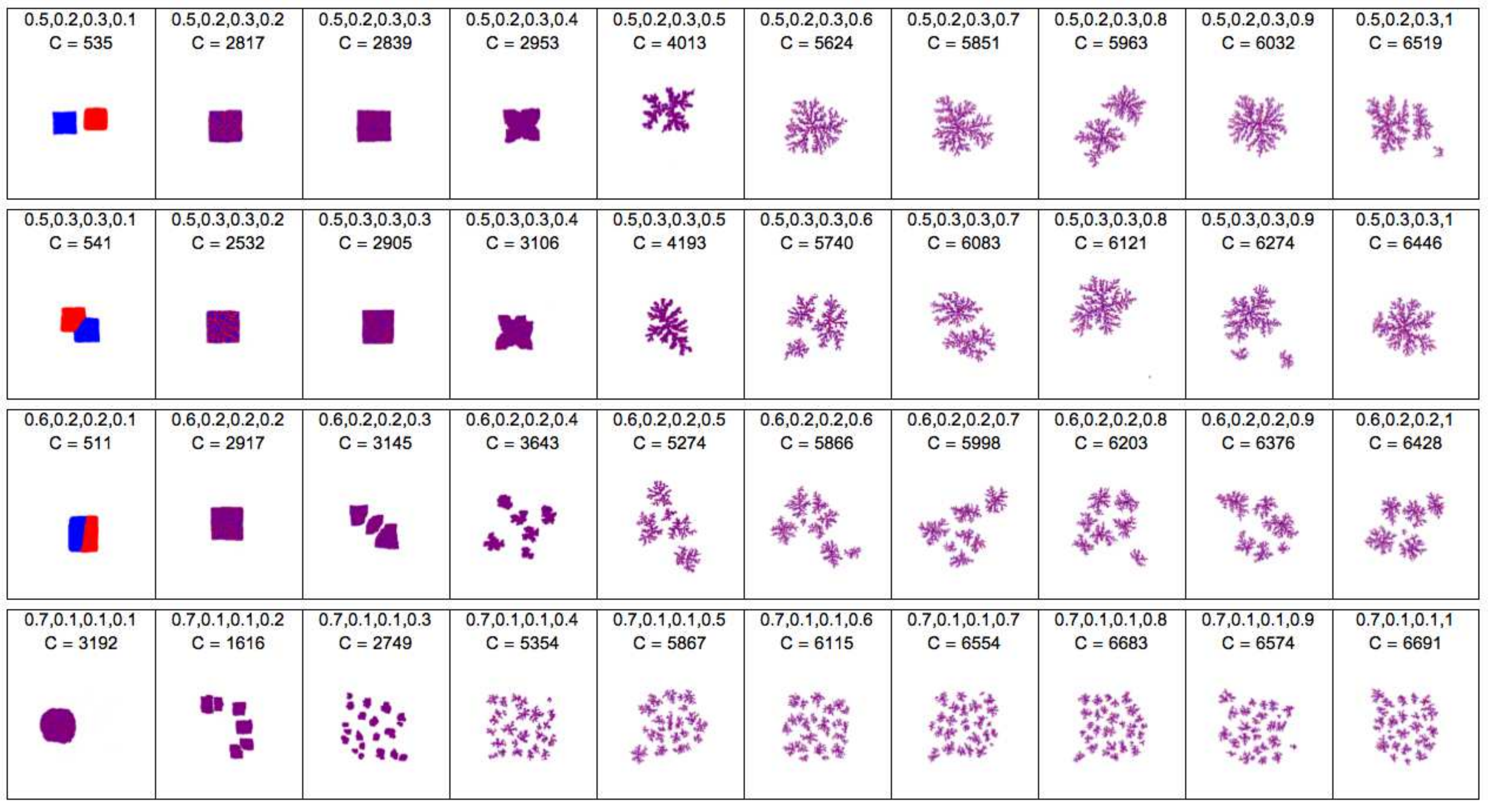}}
\caption{\label{increase}\small Simulation results are labelled with their input parameter values $E_s$, $E_{11}$, $E_{22}$, $E_{12}$ and its compression length $C$ Consecutive single parameter variation with greatest variability and fast behavioural impact.}
\end{figure*}

\subsection{Parameter discovery and orthogonality}\label{param_disc_ortho}

Although we have previously characterised and studied the simulation result space in terms of both compressibility and information content, nothing has been said about how each of the independent input parameters of the porphyrin-tiles kMC system impacts on the resulting aggregates of the self-assembly process. In synthetic biology the concept of \emph{orthogonality} (independence) of a property is of great interest given that one often wants to be able to program a system to perform a task without influencing another task. That is, a property $A$ is said to be orthogonal to $B$ if $A$ does not influence $B$. This is deeply connected with the question: {\it what input parameters change what biological traits?} Henceforth, it is important to study the impact of individual parameter changes on the behaviour of a given system.

In particular we are now interested in analysing how binding energy between molecules and substrate and the programmable structural units independently triggered the self-assembled aggregates (behaviour) captured in the simulation results. In order to conduct this analysis the simulation results were systematically grouped in such a way that, in turns, three of the associated input parameters remained fixed and the fourth one was varied in ascending order; e.g. in one group $E_{11}$, $E_{22}$, $E_{12}$ were fixed and $E_s$ varied. For each of this groups, the compressed length between consecutive simulation results was studied in order to discover any relationship in terms of information content. 

Among all the possible arrangements, the most interesting findings were observed when $E_s$, $E_{11}$, $E_{22}$ were fixed and $E_{12}$ varied. Example of this is shown in Fig. \ref{increase} where $E_s$, $E_{11}$, $E_{22}$ are set with low binding energy values and $E_{12}\in[0.1,\dots,1.0]$ runs from left to right. In general, a systematic variation in $E_{12}$ leads to an increase of $C$, except for rare cases (e.g. last row in Fig. \ref{increase} from the first to the second and from the eighth to the ninth value change). More importantly, these cases show fast phase transitions near maximal complexity after only four parameter value changes. After this point, near maximal $C$ is reached and no qualitative or quantitative change in complexity is observed.

\begin{figure*}[ht!]
\centering
\scalebox{.35}{\includegraphics{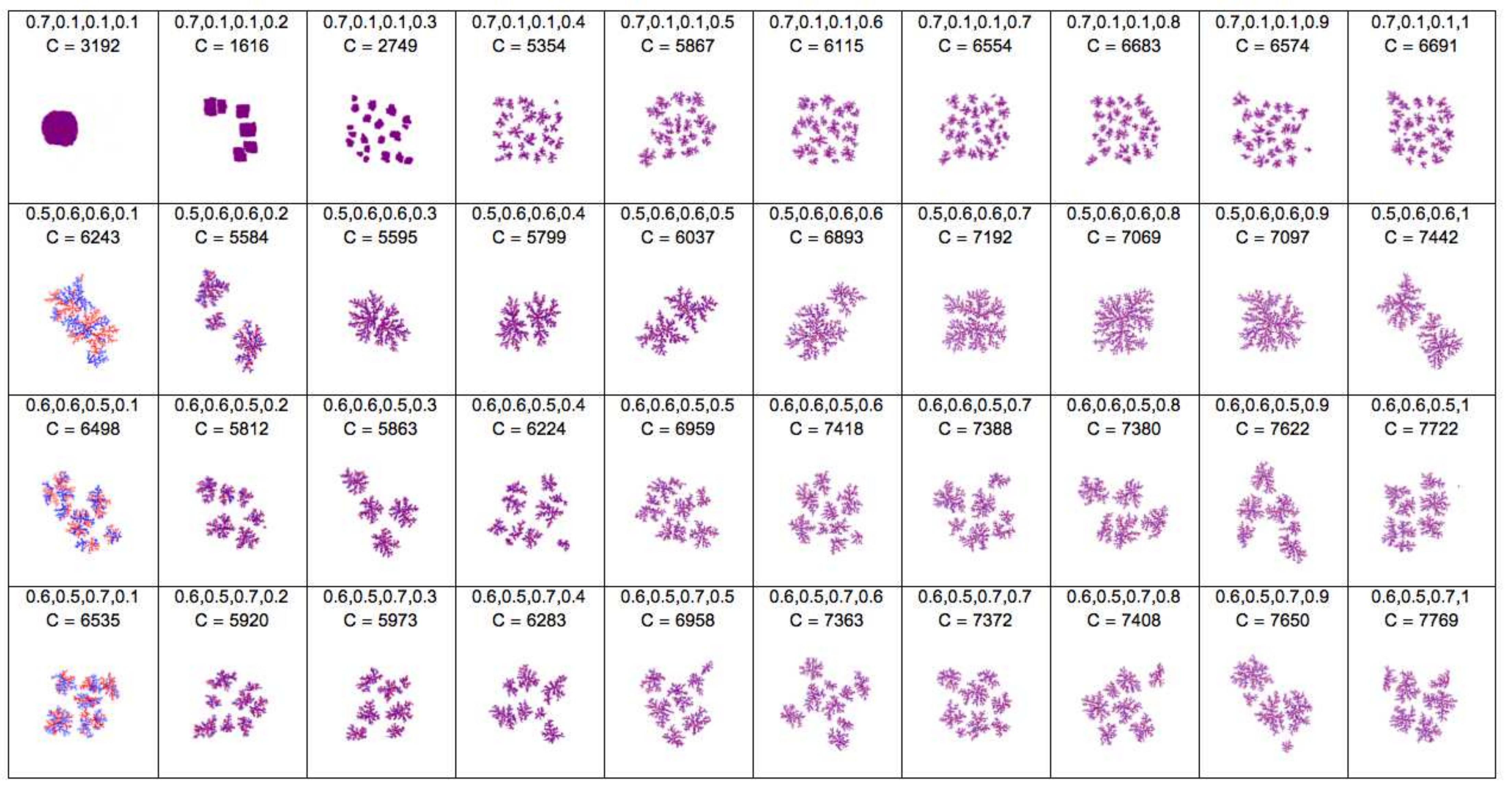}}
\caption{\label{ortho1}\small Consecutive single parameter changes that give way to reversed complexity (parameter value increases but $C$ decreases). In all cases aggregates start in a high incompressible state which is understandable in order to see reversed complexity.}
\end{figure*}

The second finding is related to reverse complexity observed when the associated compression lengths of two consecutive simulation results decrease as one of their input parameter values increases. Examples of this are found among the simulation results shown in Fig. \ref{ortho1} where each of the rows starts in a high incompressible state and then continues with a more compressible one due to variations in traits of their self-assembled aggregates. This is also a recurrent phenomenon taking place as $E_{12}$ varies towards its largest possible value.
\begin{figure*}[ht!]
\centering
\scalebox{.35}{\includegraphics{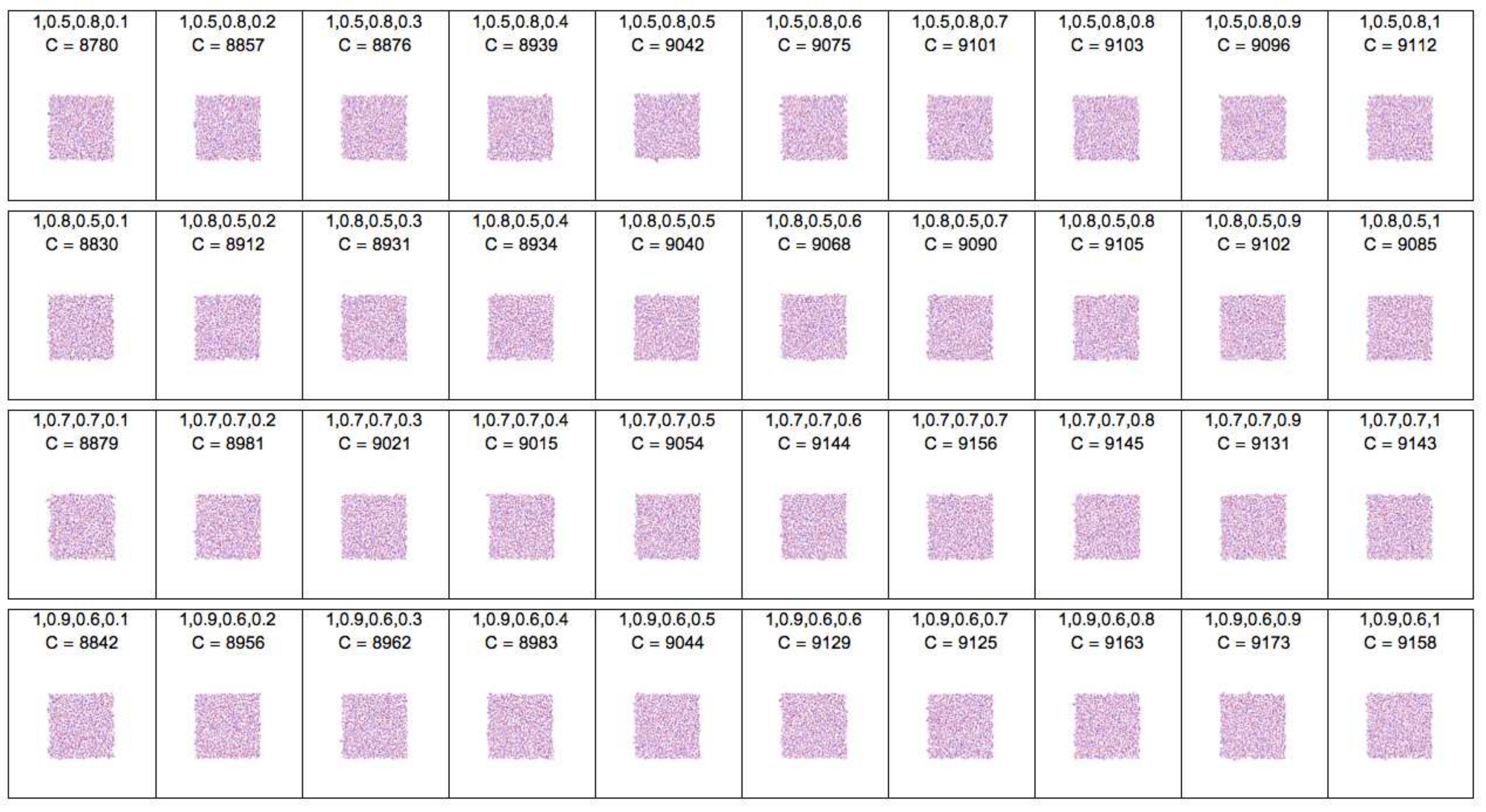}}
\caption{\label{ortho2}\small Consecutive single parameter changes with smallest qualitative impact.}
\end{figure*}

The third interesting finding is observed among simulation results employing high binding energy values as shown in Fig. \ref{ortho2}. In this case, $E_s$ is set with the largest possible value and variation occurs from left to right on input parameter $E_{12}\in[0.1,\dots,1.0]$. Contrary to what is observed in the first finding, these groups reveal that systematically changes across $E_{12}$ have a small qualitative impact on the compression length $C$ and hence little behavioural impact.

Overall, the analysis seen along Figs.~\ref{increase}, \ref{ortho1} and \ref{ortho2} shows that $E_{12}$ seems to be the most interesting input parameter in the sense that it can either produce no behavioural impact fixing $E_s$, $E_{11}$ and $E_{22}$, but also it can have the greatest impact when the values set to $E_{11}$ and $E_{22}$ are at the lowest and $E_s$ at the middle of its range. Adding a single parameter or covering a single one more thoroughly brings in a combinatorial explosion, making any systematic investigation an NP problem that requires exponential time for a linear increase in number of parameters to analyse. The numerical analysis with algorithmic complexity techniques is not necessarily computationally cheap, but it requires no human intervention, is objective and universal in mathematical terms, and does not need to be done but once in order to quantify and store what behaviour is triggered by what parameters useful for speeding up the iterative model in systems biology.

\section{Concluding Remarks}

In this work, we have introduced porphyrin-tiles which is a mathematical model that allow us to abstract and program interactions between porphyrin molecules. In particular, we focused on extensive simulations of differently programmed porphyrin molecules deposited on a solid processed $A(111)$ substrate. As a result, we have explored molecular tiling of varied morphologically complex self-assembled structures, the qualitative traits of which have been analysed in terms of complexity and information distance. 

We have performed a classification of the simulation results in terms of compressibility, as approximation to the Kolmogorov complexity, which showed that similar qualitative structural properties are arranged next to each other. From these, a phase transition diagram has been inferred showing the distribution of aggregates conformed with an intuition of increasing randomness where low (high) complexity structures are located close to (far from) the origin. In addition, a general correlation between the input parameter space and the simulation result space has been established revealing that those simulation results close to (far from) the origin of the input parameter space are highly (lowly) compressible. In addition, an important related measure of Kolmogorov complexity, i.e. normalised information distance, has been employed to investigate the similarity between self-assembled aggregates captured by the simulation results. In here we have employed NCD as distance function from which an automated classification has yielded two very well separated groups, one with low Kolmogorov complexity and another one with high Kolmogorov complexity. Given that we can numerically map the output landscape of this natural system one can think of devising precise input sequences that produce a desired targeted behaviour. The study of non-DNA based discrete molecular computation is in its infancy. This paper is but one small step into this exciting area. Several questions remain to be answered, ranging from the physical practical implementation of these tiles (work is currently being carried out in our lab) to which is the best model of computation that better describes porphyrin-based molecular computation. Is an automata-based model a good one \citep{TerLuiKras2013} or is, e.g., a Moore or Mealy machines \citep{Mealy55} or an interaction-based model better \citep{goldinwegner2006}? 

\section*{Acknowledgements}
This work is supported by EPSRC grants EPSRC EP/J004111/1 and EP/H010432/1 Evolutionary Optimisation of Self-Assembly Nano-Design (ExIStENcE). The authors acknowledge the insightful discussions on the chemistry and physics of porphyrins with Prof. N. Champness, Prof. A. Moriarty and Prof. P. Beton from the University of Nottingham.

\bibliographystyle{spbasic}

\end{document}